\documentclass[10pt, conference]{IEEEtran}
\pdfoutput=1
\IEEEoverridecommandlockouts
\usepackage{booktabs} % For formal tables
\usepackage{setspace}
\usepackage{rotating}
\usepackage{graphicx}
\usepackage{subfigure}
\usepackage{comment}
\usepackage{subfigure}
\usepackage{color}
\usepackage{dsfont}
\usepackage [english]{babel}
\usepackage [autostyle, english = american]{csquotes}
\MakeOuterQuote{"}
\mathchardef\Gamma="0100 \mathchardef\Delta="0101
\mathchardef\Theta="0102 \mathchardef\Lambda="0103
\mathchardef\Xi="0104 \mathchardef\Pi="0105
\mathchardef\Sigma="0106 \mathchardef\Upsilon="0107
\mathchardef\Phi="0108 \mathchardef\Psi="0109
\mathchardef\Omega="010A

\newcommand{\outline}[1]{}%{\textbf{#1}}

\usepackage{xspace}
\usepackage{url}
\usepackage{graphicx}
\usepackage{latexsym}
\usepackage{amsmath}
\usepackage{amssymb}
\usepackage{amsfonts}
\usepackage{psfrag}
\usepackage{subfigure}
\usepackage{wrapfig}
\usepackage{comment}
\usepackage{algorithmic}
\usepackage{alltt}
\usepackage{multirow}
\usepackage{soul}

\newcommand{\ie}{\textit{i.e.}\xspace}
\newcommand{\eg}{\textit{e.g.}\xspace}

\newcommand{\etal}{\textit{et al.}\xspace}
\algsetup{indent=1em}

\newcommand{\Comment}[1]{}

\newcommand{\iot}{IoT}

\newcommand{\iotn}{IoT\ networks}

\newcommand{\presec}{\vspace{-0.05in}}

\newcommand{\presub}{\vspace{-0.05in}}

\begin{document}
\title{IoTSense: Behavioral Fingerprinting of IoT Devices}

\author{
\IEEEauthorblockN{
Bruhadeshwar~Bezawada,
Maalvika~Bachani,
Jordan~Peterson,
Hossein~Shirazi,
Indrakshi~Ray,
Indrajit~Ray
}
\IEEEauthorblockA{Computer Science Department, Colorado State University, \\1100 Center Avenue Mall, Fort Collins, Colorado 80523 USA. \\Emails: \{bru.bezawada,~maalvika.bachani,~jordantp,~shirazi,~indrakshi.ray,~indrajit.ray\}@colostate.edu}
}

\maketitle
\begin{abstract}
The Internet-of-Things (IoT) has brought in new challenges in \textit{device
identification} --what the device is, and \textit{authentication} --is the device
the one it claims to be.
Traditionally, the authentication problem is solved by means of a cryptographic protocol.
However, the computational complexity of cryptographic protocols and/or scalability
problems related to key management, render almost all cryptography
based authentication protocols impractical for IoT.
The problem of device identification is, on the other hand, sadly neglected.
We believe that device fingerprinting can be used to solve both these
problems effectively.
In this work, we present a methodology to perform device behavioral fingerprinting that can be employed to
undertake device type identification.
A device behavior is approximated using features extracted from the network traffic of the
device.
These features are used to train a machine learning model that can be used to detect similar device types.
We validate our approach using five-fold cross validation; we report a identification rate of $86$-$99\%$ and a mean accuracy of 99\%, across all our experiments.
Our approach is successful even when a device uses encrypted communication.
Furthermore, we show preliminary results for fingerprinting device categories, \ie, identifying different device types having similar functionality.
\end{abstract}


\presec\section{Introduction}\label{sec:intro}\presec
\subsection{Motivation}\label{sec:motivation}\presub
Internet-of-Things (IoT) devices industry is rapidly growing \cite{2016report} with an ever-increasing list of manufacturers offering a myriad of smart devices targeted to enhance end-users' standard of living.
Security is an after-thought in these devices resulting in vulnerabilities \cite{2016attack} that have been successfully exploited, for instance, the notorious incident of the Mirai botnet \cite{2017mirai}.
Many security problems can be mitigated through strong identification and authentication of devices, which enables administrators to enforce appropriate security controls on a particular device.

As devices are plugged-in and removed from an \iot~ network, it is essential to identify the type of these devices and establish a behavioral baseline.
Fingerprinting IoT devices is challenging due to the large variety of devices, protocols, and control interfaces, across the devices.
An \iot~ device might respond to queries about its identity and type, which is a standard way of remotely learning about the device.
But, an untrusted device can masquerade as another device by providing false information about its identity and type.
More importantly, an untrusted or compromised \iot~device might behave contrary to its baseline behavior, \eg, connecting to other devices to disrupt their functioning or to gather network information.
Therefore, device fingerprinting IoT devices is important for achieving security in \iotn.
\subsection{Problem Description}\label{sec:problem}
\begin{figure}[hbp]
%\vspace*{-0.in}
\centering
\includegraphics[height=2.5in, width=3in]{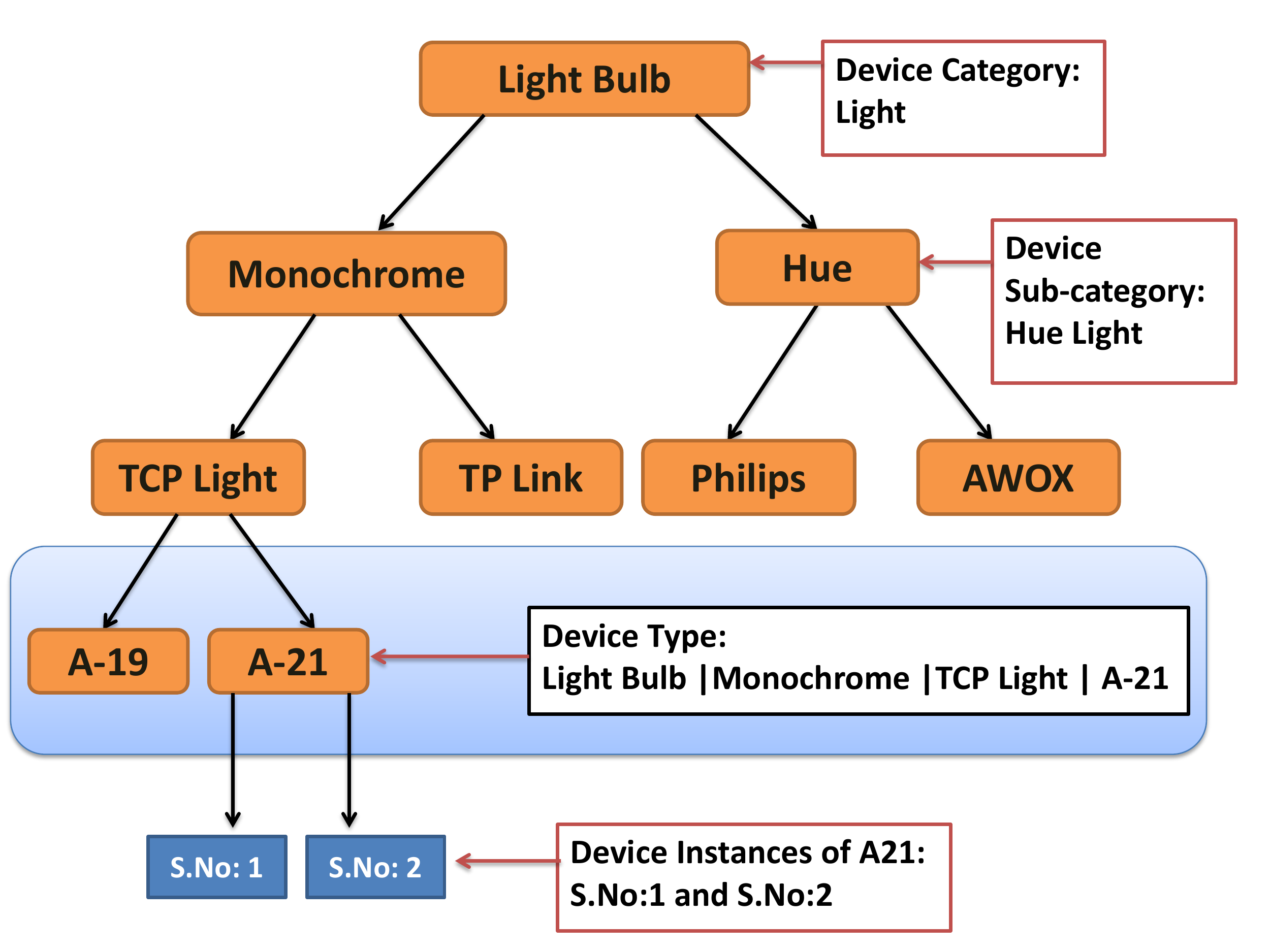}
\caption{Device Category, Type and Instance}
\label{fig:device}
\vspace*{-0.1in}
\end{figure}

An IoT device can be fingerprinted at varying levels of granularity, from a category to a specific instance, as shown in the sample ontology in Figure \ref{fig:device}.
A device category corresponds to a general grouping of devices having similar functionality, say, \eg, "Light Bulb".
This category can have further sub-divisions, like "Monochrome" and "Hue Light".
Now, a device type is a specific device model within a general device category.
For instance, in Figure \ref{fig:device}, a device type is: "Light Bulb| Monochrome| TCP Light| A21" or simply "TCP Light | A21".
Finally, a device instance is a physical device instantiation of a device type, for instance, the device type "TCP| A21", has two different bulbs with serial numbers "A21| S.No: 1" and "A21| S.No: 2".

Ideally, a security administrator would like the capability of establishing unforgeable identities, \ie, authenticating, two device instances, say, "A21| S.No: 1" and "A21| S.No: 2", which are of the same device type.
Towards this goal, device type identification is a critical first step.
\noindent{\bf Problem Statement.    }   \   We describe the problem of fingerprinting an \iot~ device type as that of identifying the device type from a sample network activity of the device.
We refer to the sample network activity of a device, $D_i$, as its \textit{fingerprint,} $F_i$.
The collection of all possible network activities of a $D_i$ constitutes the behavioral profile, $\langle\mathbb{B}_i, D_i\rangle$, of the device.
From this discussion, the problem statement we address is as follows.

Given a collection of previously recorded behavioral profiles, $\mathbf{B}=\{\langle \mathbb{B}_1,~D_1\rangle,~\langle\mathbb{B}_2,~D_2\rangle,~\cdots,~\langle \mathbb{B}_n,~D_n\rangle\}$ of $n$ devices and the available fingerprint $F_t$ of a target device $D_t$, to correctly predict $\langle\mathbb{B}_t,~D_t\rangle$ where $\mathbb{B}_t (\ni F_t)$ is the corresponding behavioral profile of $D_t$.
\subsection{State of Current Research}
IoT device type fingerprinting research is in early stages due to the evolving nature of the IoT industry.
%
%We describe the issues not addressed by the current research in this area.
%

Miettinen \etal in \cite{sadeghi} described IoT Sentinel, a framework for device fingerprinting and securing IoT networks.
Their work focused on machine learning techniques for fingerprinting a device when it first registers on a network.
However, their work does not analyze the behavior of a device as described the problem statement.
Our work is complementary to theirs, it can be used along with their approach to fingerprint devices and provide stronger security.

Siby \etal described IoTScanner in \cite{2017siby}, an architecture that passively observes network traffic at the link layer, and analyzes this traffic using frame header information during specific observation time windows.
This work is more concerned with discerning the distinct devices and their presence based on the traffic patterns observed during the traffic capture time window.
A shortcoming of this approach is that two identical device types could be classified as two different device types due to the variations in traffic generated during traffic capture time window.
Device fingerprinting, wireless and wired, has received considerable attention in the research community.
General device fingerprinting has been described in \cite{1997martin, 2003lippmann, 2005kohno}, which have explored several techniques ranging from packet header features to physical features such as clock-skews.
Wireless device finger printing techniques have been discussed in \cite{2006jason, 2007pang, 2009james, 2010chrisil,  2016kurtz}.
These works explored the device type identification by exploring the implementation differences of a common protocol such as SIP, across similar devices.
However, IoT devices use numerous protocols and it would be nearly impossible to attempt such analysis on a per protocol and per device basis.
Physical layer based device fingerprinting has received considerable attention \cite{2008wireless,2010jana,2015beyah,2016formby,2016van} where the focus is on analyzing the physical aspects of devices to fingerprint them.
All these works focused on general wireless devices and their applicability to IoT devices is an open question.
\subsection{Proposed Approach, Technical Challenges and Solutions}\label{sec:approach}
Our fingerprinting approach generates a behavioral profile that quantifies the behavior of a device type.
Behavioral fingerprinting is quite valuable since it allows us to monitor the device behavior throughout its life time.
If there are deviations from the device's initial behavior, due to some malicious activity, we can detect such activity by periodically observing and validating against a behavioral profile.
To generate such a behavioral profile, our approach is to model the behavior of the device \textit{approximately} as a collection of protocols used, and the set of observed command and response sequences.
We collect the network traffic that is flowing into and out of the device and extract features of interest are indicators of a device behavior.
Finally, we aggregate the features using a statistical model and use it as a reference for identifying the device.
To identify a target device, we observe a few packets from the device and compare it against the previously recorded behavioral profiles.
%
%\subsection{Technical Challenges and Solutions}
%

The first technical challenge in our proposed approach is to be able to observe all possible protocol interactions and command-responses of the device.
Since it may not be possible to observe all possible interactions, our approach can only \textit{approximately} model the device behavior.
To solve this challenge, in the laboratory setting, we use the controlling smart-phone app to interact with the device to extract the command-response sequences and this coupled as well as passively observed the device.
Our approach is essentially simulating the passive observation of network traffic where the observer could be observing the traffic flows of a new device to build the behavioral profile.

% %%% Challenge:
The second challenge is that the types of interesting behavioral features are non-trivial to determine.
Therefore, use available features like packet header feature and payload based features for this purpose.
For packet header based features, we extract the device specific features such as the protocols used and communication patterns of the device during the observational period.
Our choice of payload features, coupled with some preliminary results, show that the our approach can work on encrypted traffic as well.
%

% #We analyze the data using a statistical model.
The next challenge in our proposed approach is that statistical models can be very difficult to generate on multi-variate data.
Towards this, we apply general purpose machine learning tools as machine learning classifiers are very good at learning local features of interest in a collection of data.
Since, we are modeling the behavior of a device as a collection of individual fingerprints, the machine learning classifiers are most suitable for our approach.
Typically, one distinguishing feature is sufficient to classify a given device type against several other device types and this depends entirely on the robustness of the machine learning model.
%

% Challenge:
The final and most important challenge is: \textit{How much of the target device data needs to be observed before the fingerprint can be matched against a stored behavioral profile?}
Ideally, a small number of packets allows a fingerprinter to be able to keep track of the devices periodically and observe
any deviations from its behavior.
%
%But how to determine how many packets are required?
Our approach therefore attempts to create a short fingerprint with a few packets being sufficient to match the fingerprint.
Therefore, the problem is to determine, on an average, the number of device packets that encapsulate one or more behavioral features of a target device.
To solve this problem, we determined the average number of packets based on the assumption that an individual protocol interaction session of a device encapsulates one or more behavioral features of the device.
This is consistent with our model for the behavioral profile of a device, which maps the device behavior as collection of protocol interaction sessions.
Using experimental analysis, we establish the approximate number of packets required for fingerprinting a device.

\noindent{\bf Key Contributions.}       \  (a) For the first time, we describe a practical approach for behavioral fingerprinting of IoT devices using machine learning.
(b) Our fingerprinting analysis shows that with a small number of packets we can fingerprint a device quite accurately.
%
%We provide detailed analysis that led to these results.
%
(c) We demonstrate that certain features like TCP window size, entropy and payload lengths are very specific to device types and are statistically significant in fingerprinting.
(d) We demonstrate machine learning model robustness using three different types of experiments on $14$ different device types and $9$ different device categories.
The first experiment examines the identification of device types based on a five-fold cross-validation and, we report a mean identification rate in the range of $93$-$99\%$, and a mean accuracy of $99\%$.
The second experiment examined the identification of device categories, and using five-fold cross-validation, we report a mean identification rate of $91$-$99\%$, of identifying a device to its proper category.
This result is the first such success reported in this problem domain and shows that we can create behavioral profiles for classes of devices.
The final experiment examined identification of different device instances of same device type, and achieved an excellent mean identification rate of, $99.7$-$100\%$.
(e) Finally, with our existing device set, we show preliminary results that our approach is successful even when the device uses encryption for some of the communication.
%

%\input{system}
%
% Discuss different fingerprinting approaches as written up earlier including IoTScanner and IoTSentinel
%
%
\section{Related Work}\label{sec:related}
In \cite{2006jason}, Franklin \etal describe a passive fingerprinting method for identifying  the different types of 802.11 wireless device driver implementations on clients.
The  authors explore the statistical relationship of the active channel scanning strategy in a particular device driver implementation.
The lack of a standard for the scanning strategy resulted in such differences and allowed the authors to distinguish between the devices.
This technique is useful for identifying the type of device driver implementation but not of the type of device.
For instance, a manufacturer might reuse the same device driver  implementation across several device types.

In \cite{2008wireless}, Vladimir \etal  developed a radiometric approach based on imperfections in analog components for fingerprinting network interface cards (NICs).
Such variations result in imperfect emissions when compared with the theoretical emissions and manifest in the modulation of the transmitted signals of the device.
They used machine learning approaches to perform the fingerprinting.
However, this work relies on the availability of the frames, physical layer transmission, from the given device.
This may not be for an IoT network as the devices are spread overa an area and might be interconnected via different switches and middle-boxes.

In \cite{2009james},  Fran{\c{c}}ois \etal describe approaches to fingerprint devices based on the usage of a common protocol.
Broadly, this work focuses on distinguishing various implementations of the same protocol.
This work describes interesting techniques to parse protocols and provides an approach for representing and anlaysing the behavior of a given protocol.
However, IoT devices speak a variety of protocols, which makes it difficult to apply these techniques.

In \cite{2010james},  Fran{\c{c}}ois \etal  describe a protocol grammar based approach for fingerprinting.
Similar to our work, they characterize a device based on the set of messages emitted.
A message is represented using the protocol grammar syntax.
To classify a given device, the messages emitted by the device are compared with syntactic trees of the stored fingerprints and depending on a similarity metric, the device label is assigned.
However, this approach is again specific to protocols that are well known and whose grammar rules are available.

In \cite{2010gao}, Gao \etal developed a wavelet analysis technique to fingerprint wireless access points based on frame inter-arrival time deltas.
This technique can be seen as a black-box approach.
However, this approach relies on the fingerprinter being in or near the range of the access point to gather sensitive time information and needs the access point to route data to the fingerprinter.
The approach does not apply to IoT devices as these devices are usually end-points and do not forward data to other devices.

In \cite{2015beyah} Radhakrishnan \etal described GTID for device-type identification on general purpose devices like smartphones, laptops and tablet PCs.
Their work relies on the inter-arrival times of different packets to extract the relevant features specific to a particular application like Skype.
However, IoT devices are usually very conservative in terms of traffic generation and do not generate much traffic, as we have observed in our laboratory setting.
Applying these techniques to IoT networks will require non-trivial modifications to the original set of algorithms.
In contrast, our work extracts the behavior of an IoT device on whatever traffic is made available.

In this problem space, IoTSentinel \cite{sadeghi} by Miettinen \etal and IoTScanner \cite{2017siby} by Siby \etal are the currently known solution frameworks.
IoTSentinel focuses on device-type identification at the time of device registration into a network.
This approach uses packet header based features to identify a particular device type.
IoTSentinel reports a mean identification rate of $50$-$100\%$, whereas our approach reports a mean identification rate of $93$-$99\%$.
Our approach complements IoTSentinel, as our approach can periodically cross-verify the device fingerprints established at registration time.
IoTScanner \cite{2017siby}, is similar to GTID \cite{2015beyah}, in that they identify devices by visualizing the MAC (medium access control) layer traffic of the devices.
This approach is useful for network mapping at a high level, but performing this analysis periodically can be cumbersome.
In contrast, our approach can re-verify a fingerprint of a device with a short signature of only $5$ packets.
\section{Behavioral Fingerprinting Model}\label{sec:fingerprint}
In this section, we describe the building blocks of our behavioral model of an IoT device.
First, we describe the \textit{static } behavioral model of an IoT device in terms of the protocols used by a given device.
Second, we describe the \textit{dynamic} behavioral model of IoT device in terms of the session interactions of the device.
\subsection{Static Behavioral Model}\label{sec:static}
IoT devices use different protocols at different stages of their operation such as may include a subset of ARP, SSL, LLC, EAPOL, HTTP, MDNS and DNS.
Therefore, the list of protocols used by an IoT device is a good indicator of the device behavior.
In \cite{sadeghi}, the authors used this notion to capture the device behavior at registration time.
However, the list of protocols used by the device provides only a \textit{static} view of the device's operations.
In addition to this, further modeling is required to completely understand the  dynamic nature of the device's behavior.
\subsection{Dynamic Behavioral Model}\label{sec:dynamic}
Our modeling of the dynamics of an IoT device is based on the notion that an IoT device has several distinct command-response sequences.
We call each of these command response sequences as a \emph{session}.
For instance, let us consider that a device responds to (or sends) the following types of control messages: $C_1,~C_2,\ldots,~C_n$ and that the responses for each of these messages are (not necessarily in that order): $R_1,~R_2,\ldots,~R_m$.
A typical protocol interaction can be as follows: $C_1\rightarrow R_1 \rightarrow C_2 \rightarrow R_3 \rightarrow C_1 \rightarrow R_1$.
Therefore, the device's behavior can be viewed as a collection of these sequences.
However, the challenge is in estimating the average number of packets that are part of any given session.

One way to to estimate this average is by considering the limited scope of IoT devices, which usually have short sessions consisting of $2$ to $10$ packets.
Given this intuition, the average number of packets per session, across $5$ such devices, is given by: $\frac{(2+4+6+8+10)}{5}=6$ packets.
To check this theoretical result, we used the data from our experiments to count the number of sessions and the packets per session across a variety of devices.
To count a session, we considered all packets exchanged with the same destination and source ports and we show the results in Table \ref{tbl:sessions}.
For this sample set of data, the average number of packets per session is: $\frac{(3.89+6.18+4.41+5.13+9.48+9.77+8.79)}{7}=6.8$, which is very close to our theoretical estimate.
The summary of this result is that, to fingerprint a given device we need to capture $6\pm2$ packets for any given device.
A limitation is that there are certain devices that act as conduits, such as Philips Hub, or devices that do not have sessions, and cannot be modeled by this approach.
However, using experimental analysis, we demonstrate that such devices can be still be fingerprinted using our approach due to their limited behavioral profiles.
\begin{table}[hbp]
\footnotesize
%\centering
\caption{Average Packets Per Session}
\vspace*{-0.1in}
\label{tbl:sessions}
\begin{tabular}{|l|l|l|l|}
\hline
\multicolumn{1}{|l|}{\textbf{Device}} & \textbf{Total Sessions' Packets} & \textbf{Sessions } &\textbf{Packets/Session} \\\hline
AWOX Speaker                           & 12755	&3274 & 3.89 \\\hline
D-Link Camera                           &8600	& 1390&6.18 \\\hline
MUSAIC Speaker                     & 1346	&305&4.41  \\\hline
OMNA Camera                                     & 8253&	1608&5.13\\\hline
TP Link Light                            & 1660&	175&9.48  \\\hline
WEMO Outlet                        &1994&204&9.77 \\\hline
WINK Hub                            & 739&	84 & 8.79 \\\hline
\end{tabular}
\end{table}
Based on the models described in this section, we describe the feature selection for the machine learning models that will be used to create the behavioral profile of the IoT devices.
\begin{figure*}[htbp]
%\centering
%\hspace*{-0.2in}
\parbox{7in}{
\subfigure[Entropy]{\label{fig:ent}
  \includegraphics[width=2.3in, height=2in]{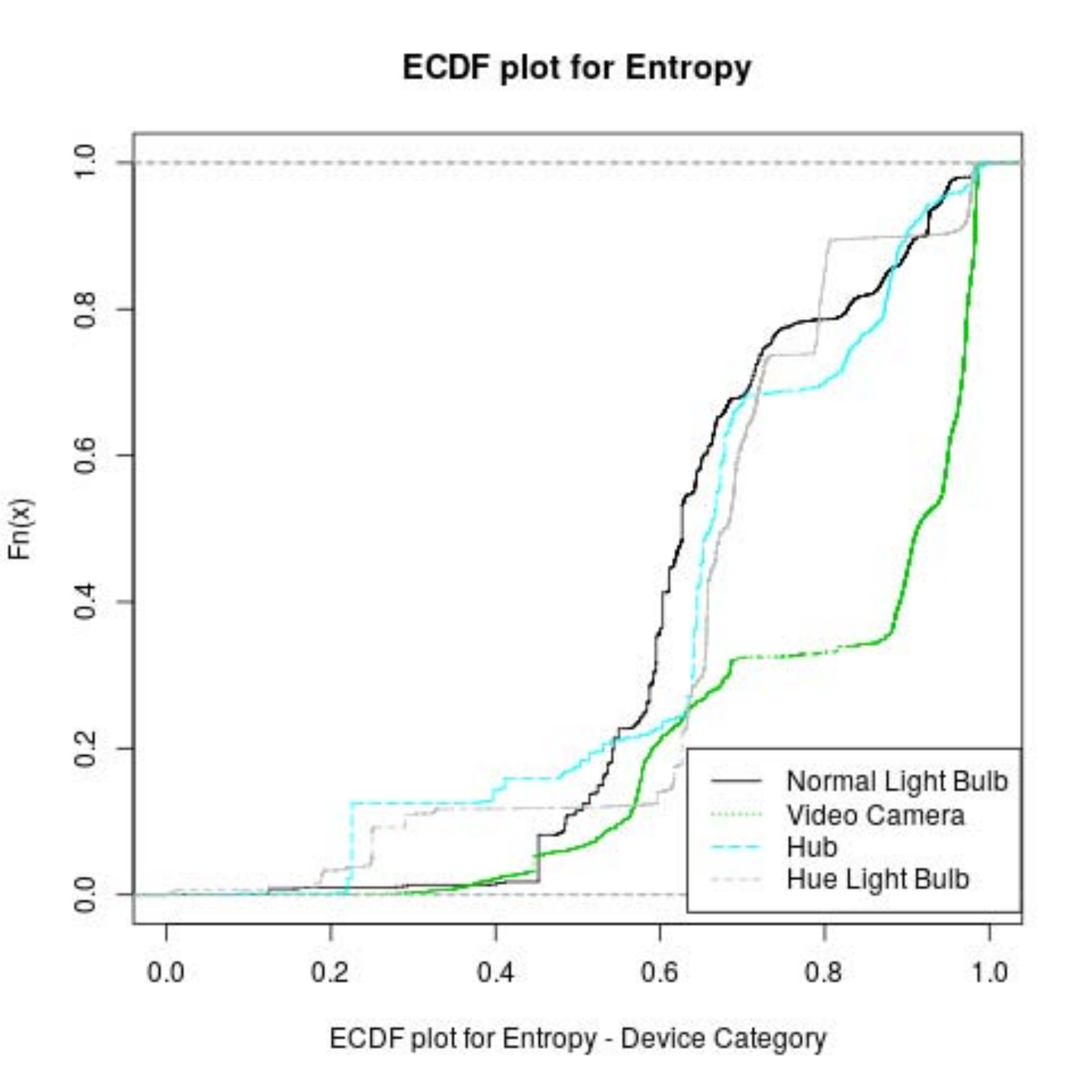}
  }  %\hspace*{-0.1in}
\subfigure[Payload Length]{\label{fig:payl}
  \includegraphics[width=2.3in, height=2in]{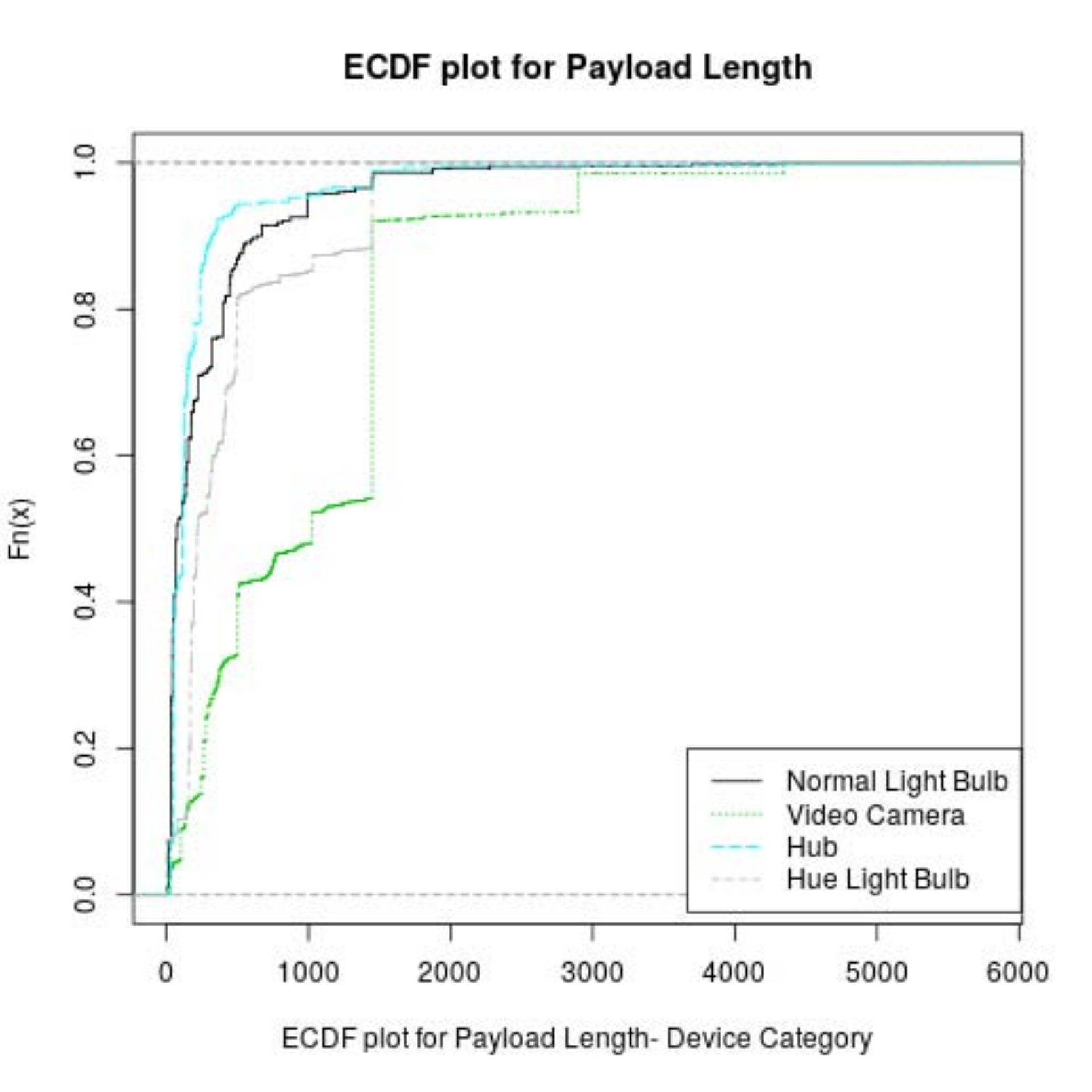}
  }
\subfigure[TCP Window Size]{\label{fig:tcpw}
  \includegraphics[width=2.3in, height=2in]{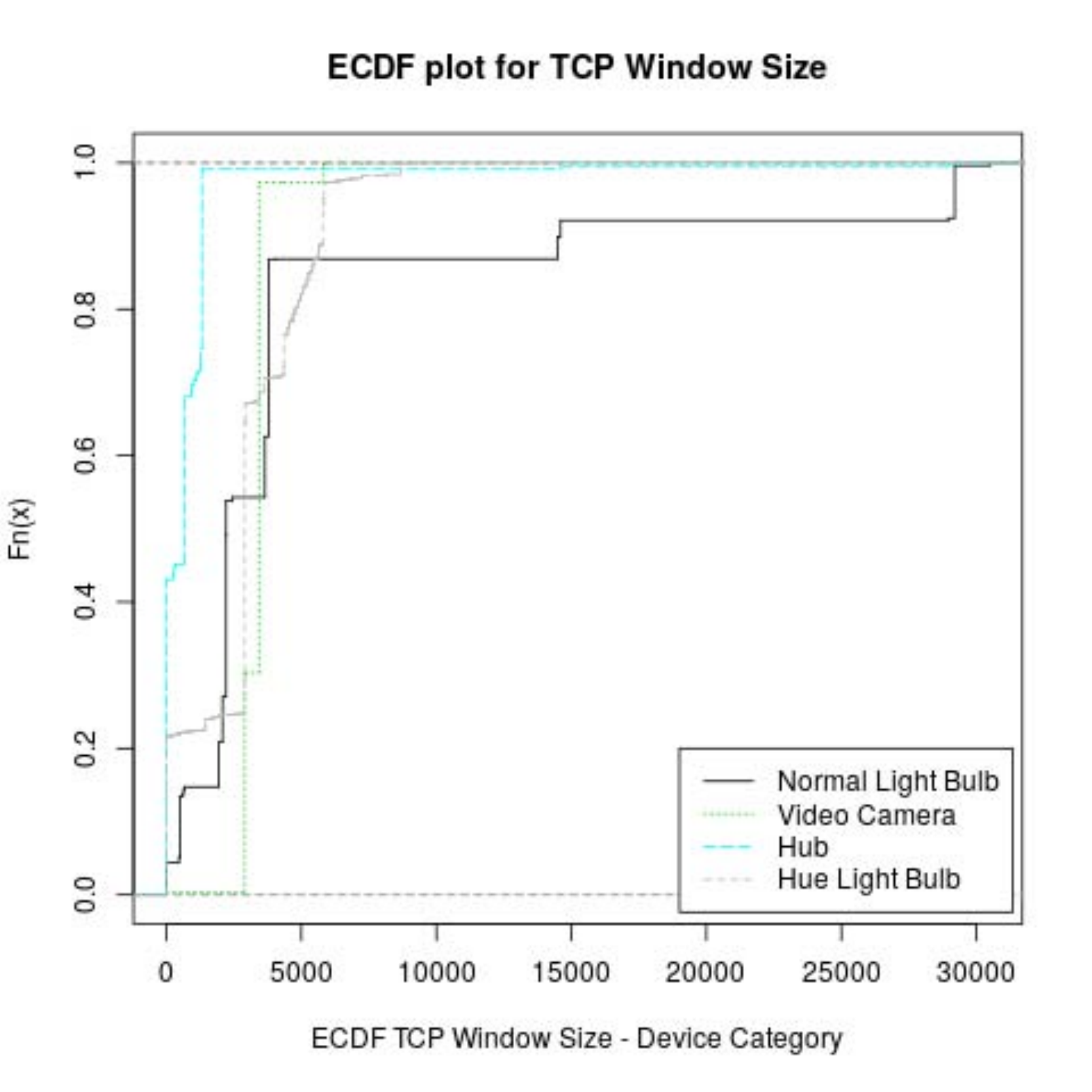}
  }
%\vspace*{-0.2in}
\caption{ECDF of Payload Based Features}\label{fig:ecdf}
}
\vspace*{-0.1in}
\end{figure*}
\section{Machine Learning Features for Behavioral Profiling}\label{sec:features}
We use two available types of features from the network packets: packet header features and payload based features.
Broadly speaking, the packet header features are useful in quantifying the static behavioral model of the device, and the payload based features are useful in quantifying the dynamic behavioral model of the device.
\subsection{Packet Header Features}\label{sec:packet}
\begin{table*}[htbp]
\small
\centering
\caption{Packet Header Features}
\vspace*{-0.1in}
\label{tbl:features}
 \begin{tabular}{|l|l|}
 \hline
   {\bf Protocol Layer/Type} & {\bf Features} \\
   \hline
   % after \\: \hline or \cline{col1-col2} \cline{col3-col4} ...
   Network  & IP/ICMP/ICMPv6/EAPoL \\ \hline
   Transport  & TCP/UDP \\ \hline
   Application  & HTTP/HTTPS/DHCP/BOOTP/SSDP/DNS/MDNS/NTP \\ \hline
   IP Options & \emph{Padding}/\emph{Router Alert} \\
%   Payload Length & Size(int) \\
 %  Packet entropy & Value (float)\\
   \hline
 \end{tabular}
\end{table*}
For the static behavioral model, we use a subset of the features, shown in Table \ref{tbl:features}, from those outlined by Miettinen \etal in \cite{sadeghi}.
Essentially, these features are extracted from the packet headers of the traffic data from the device.
These features are binary, \ie they have values of $0$ or $1$ for the absence or presence of a feature, respectively.
Note that, unlike the work in \cite{sadeghi}, we do not consider network specific features like IP addresses, source or destination ports and so on, as these features are not dependent on the device behavior.
\subsection{Payload Based Features}
Primarily, we consider the use of three important features: entropy of payload, TCP payload length and TCP window size.
To validate the intuition behind each feature, we tested the empirical cumulative distribution function (ECDF) of the feature for four different types of devices.
The  ECDF of a real-valued random variable $X$, or just distribution function of $X$, evaluated at x, is the probability that $X$ will take a value less than or equal to x.
For $x$-axis distribution, we used the feature values in the dataset, and for $y$-axis, we used the probability that feature
value will take values less than or equal to x.

\noindent{\emph{Entropy.    }   }   \   The entropy of the payload is basically indicative of the information content inside a packet.
To calculate Shannon entropy of a sequence of $m$ bytes with a symbol length of $8$-bits or $1$ byte, the following formula is used:
$$h_m=-\sum\limits_{i=1}^{256} p_i \log_{256} p_i$$
where $p_i$ is the probability of the occurrence of byte value $i$ in the $m$ bytes, \ie, $p_i=\frac{count(i)}{m}$.
Using the analysis described by Khakpour \etal in \cite{2013amir}, if a packet is carrying plain-text then the entropy of the payload is less and, if the packet is carrying audio data, then the entropy will be high.
In using entropy as a feature, we are only focusing on the nature of the data and not on the data itself.
We performed a statistical analysis of this feature across a few devices and show the result in Figure \ref{fig:ent}.\\

\noindent{\emph {TCP Payload Length. }  } \
This is the length of the payload carried inside a TCP message, in other words, this is indicative of the length of the messages sent by a given device.
This is a very device specific feature and shows significant variation from device to device as shown in Figure \ref{fig:payl}.
Even if the messages are encrypted, for instance, the video camera feeds, the underlying block-cipher and padding result in deterministic patterns in the command and response message payload lengths.
Typically, we observed that most control message exchanged by the device from the smart phone or over the local area network are in plain-text.
For a given protocol interaction for a device these are unlikely to change and therefore, are a good indicator of the device behavior.\\
\noindent{\emph {TCP Window size. }  } \ This feature has been suggested by Alvin \etal in \cite{1997martin} as method to fingerprint general purpose devices.
The intuition behind this feature is that the TCP window size depends on the memory of the IoT device and the speed of its processing.
Small constrained devices, such as light bulbs, typically tend to have small window sizes and more powerful devices, such as video cameras, have variable and larger window sizes.
Figure \ref{fig:tcpw} shows the variation of the TCP window sizes across different categories of devices and show the variability of this feature among these devices.
Such variability is the key factor for effective machine learning based classification.
\presec\subsection{Behavioral Profile and Fingerprint}\label{sec:behavior}
Based on the discussion so far, we now define the structure of a device type fingerprint.
From Section \ref{sec:problem}, the behaviorial profile of a device is defined as a collection of various \textit{fingerprints}.
Based on the analysis done in Section \ref{sec:dynamic}, the number of messages in a session contribute to the fingerprint of that session, which is $6 \pm 2$.
We choose five packets as the number of session packets whose features correspond to a fingerprint of the device.
This implies that any given set of five session packets will represent a fingerprint of the device and should be sufficient to identify the device.
For each of the five packets we extract $20$ features, \ie, the $17$ packet header features and the $3$ payload based features,  and group them together, to give us a feature vector of $100$ features.
We consider consecutive packets, \ie, $p_i \rightarrow p_{i+1} \rightarrow p_{i+2} \rightarrow p_{i+3} \rightarrow p_{i+4}$, to generate a single feature vector, as the sequence of the packets is important to capture the session semantics.
This feature vector represents the fingerprint of the device with respect to the five chosen packets.
Now, to create a behavioral profile from the network traffic captured from a device, we group the packets into groups of five and generate the feature vectors.
The set of all such feature vectors corresponds to the observed behavioral profile of the device.
These feature vectors can then be used to train a machine learning classifier that will be able to predict the device type when presented with a target feature vector of the same device type.
\presec\section{Performance Evaluation}\label{sec:performance}
In this section, we describe our experimental setup and the various devices on which the fingerprinting tests were carried out.
We report several interesting results with different variations of features.
\presub\subsection{Experimental Setup and Data Sets}\label{sec:exp}
We tested our approach on the latest home IoT devices, listed in Table \ref{tbl:devices}, available in the market.
The \textit{device label} corresponds to the unique identifier given to this device type.
The \textit{category} corresponds to the general category under which one or more devices are grouped, \eg, AWOX light and iView light are grouped together.
The \textit{connectivity} refers to the physical layer connectivity supported by these devices.
\begin{table*}[htbp]
\small
\centering
\caption{Device Descriptions}
\vspace*{-0.1in}
\label{tbl:devices}
\begin{tabular}{|l|l|l|l|}
\hline
\multicolumn{1}{|l|}{\textbf{Device Label: Device}} & \textbf{Model} & \textbf{Category } & \textbf{Connectivity} \\\hline
1: TCP Light                            & GL30002-TP &Light & Wi-Fi  \\\hline
2: AWOX Light                          &SLCW13-14:D4:41	& Hue light & Wi-Fi \\\hline
3: MUSAIC Music Speaker                     & MP10	&Music Player &Wi-Fi, Ethernet  \\\hline
4: D-Link Camera                               &DCS-932L& 	Camera &Wi-Fi, Ethernet \\\hline
5: iDevice Socket                            & IDEV0002&Socket& Wi-Fi, Bluetooth   \\\hline
6: iView Light                        &R60&Hue light& Wi-Fi  \\\hline
7: Lutron Hub                            & L-BDG2&	Hub & Wi-Fi   \\\hline
8: Netatmo Climate                        &Home Coach&Climate Control&Wi-Fi  \\\hline
9: Omna Camera                      &DSH-C310&Camera&Wi-Fi  \\\hline
10: Philips Hue Light                       & Hue 2.1&Light&Wi-Fi   \\\hline
11: TPLink Light                        &Lb100&Hue Light&Wi-Fi  \\\hline
12: WEMO Outlet                        &Insight&Outlet &Wi-Fi  \\\hline
13: Wink Hub                        &2&Light& Wi-Fi \\\hline
14: SmartThings Hub                       & &Hub&Wi-Fi\\\hline
\end{tabular}
\end{table*}
\begin{figure}[hbp]
\vspace*{-0.1in}
\centering
\includegraphics[height=2.3in, width=2.3in]{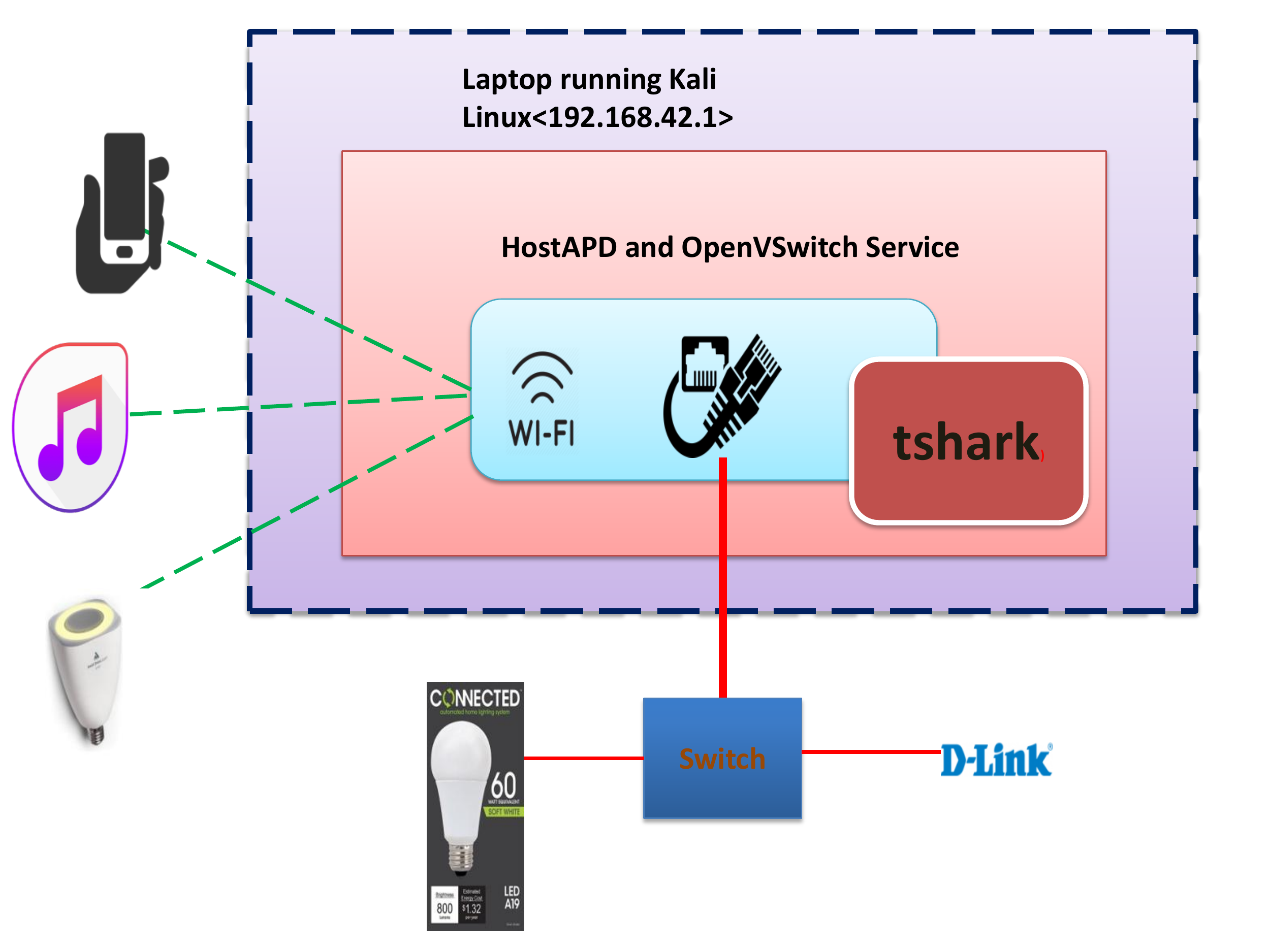}
\caption{Packet Capture Setup}
\label{fig:capture}
\vspace*{-0.1in}
\end{figure}

To enable data capture from these devices, we constructed a software bridge setup as shown in Figure \ref{fig:capture}, using a general purpose laptop running Kali Linux on an Intel$^\circledR$ processor with $8$ GB RAM.
This setup allowed us to capture all traffic, including the traffic passing through the network switch.
\subsection{Data Set Collection}
\begin{table*}[htbp]
\small
\centering
\caption{Data Per Device-type}
%\vspace*{-0.1in}
\label{tbl:data}
\begin{tabular}{|l|c|}
\hline
\multicolumn{1}{|l|}{\textbf{Device Label: Device}} & \textbf{Data Instances} \\\hline
1: TCP Light                            &   1151  \\\hline
2: AWOX Light                          &  2000\\\hline
3: MUSAIC Music Speaker                     &  1003 \\\hline
4: D-Link Camera                               & 1991 \\\hline
5: iDevice Socket                            &   415  \\\hline
6: iView Light                        &  571 \\\hline
7: Lutron Hub                            &  108  \\\hline
8: Netatmo Climate                        &  70  \\\hline
9: Omna Camera                      &   1072\\\hline
10: Philips Hue Light                       & 986  \\\hline
11: TPLink Light                        & 519  \\\hline
12: WEMO Outlet                        &  592 \\\hline
13: Wink Hub                        & 286\\\hline
14: SmartThings Hub                       &103 \\\hline
\end{tabular}
\end{table*}
To collect the necessary data sets for training the machine learning classifiers, we emulated the normal usage of a device, \ie, by controlling it with a smart phone app or through a web interface.
Our method for data set collection is as follows.
First, the device is booted up and allowed to perform any initial configuration or firmware updates.
Second, when the device is in steady state, we contacted the device through its smart app and started interacting with the device.
We also allowed periods of idle time for the device to perform some communication without user intervention.
Depending on the device activity, we captured $1000$ to $10000$ packets of network traffic from each device.
The various device operations are described in Table \ref{tbl:operations}.
In a real network environment, our approach works in a passive manner by observing all traffic and generating the corresponding behavioral profile of the device.
\begin{table*}[htbp]
\small
\centering
\caption{Device Operations}
%\vspace*{-0.1in}
\label{tbl:operations}
\begin{tabular}{|l|l|}
\hline
\multicolumn{1}{|l|}{\textbf{Device Label: Device}} & \textbf{Mode of Operation} \\\hline
1: TCP Light                            & Connects through a Hub  \\\hline
2: AWOX Light                          &  Connects with a mobile app\\\hline
3: MUSAIC Music Speaker                     &  Connects with a mobile app \\\hline
4: D-Link Camera                               & Connects through laptop \\\hline
5: iDevice Socket                            &   Connects with a mobile app  \\\hline
6: iView Light                        &  Connects with a mobile app \\\hline
7: Lutron Hub                            &  Connects with a mobile app  \\\hline
8: Netatmo Climate                        &  Connects with a mobile app  \\\hline
9: Omna Camera                      &   Connects with a mobile app \\\hline
10: Philips Hue Light                       & Connect through a Hub  \\\hline
11: TPLink Light                        & Connects with a mobile app  \\\hline
12: WEMO Outlet                        &  Connects with a mobile app \\\hline
13: Wink Hub                        & Connects with a mobile app \\\hline
14: SmartThings Hub                       &Connects with a mobile app \\\hline
\end{tabular}
\end{table*}

\begin{figure*}[htbp]
%\centering
%\hspace*{-0.2in}
\parbox{6in}{
\subfigure[20 Features]{\label{fig:20-ci}
  \includegraphics[width=3in, height=2in]{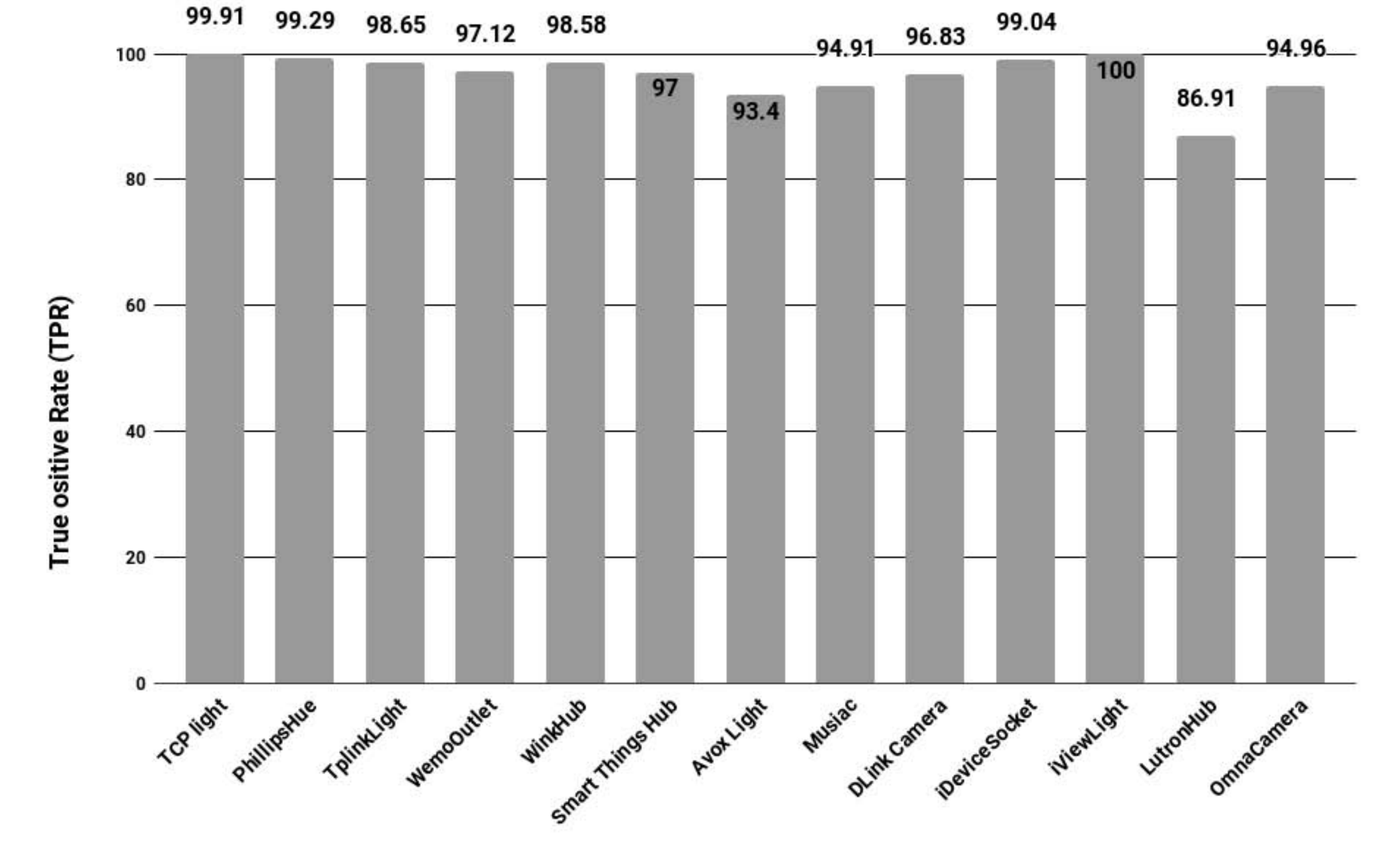}
  }
\hspace*{-0.1in}
\subfigure[19 Features]{\label{fig:19-ci}
  \includegraphics[width=3in, height=2in]{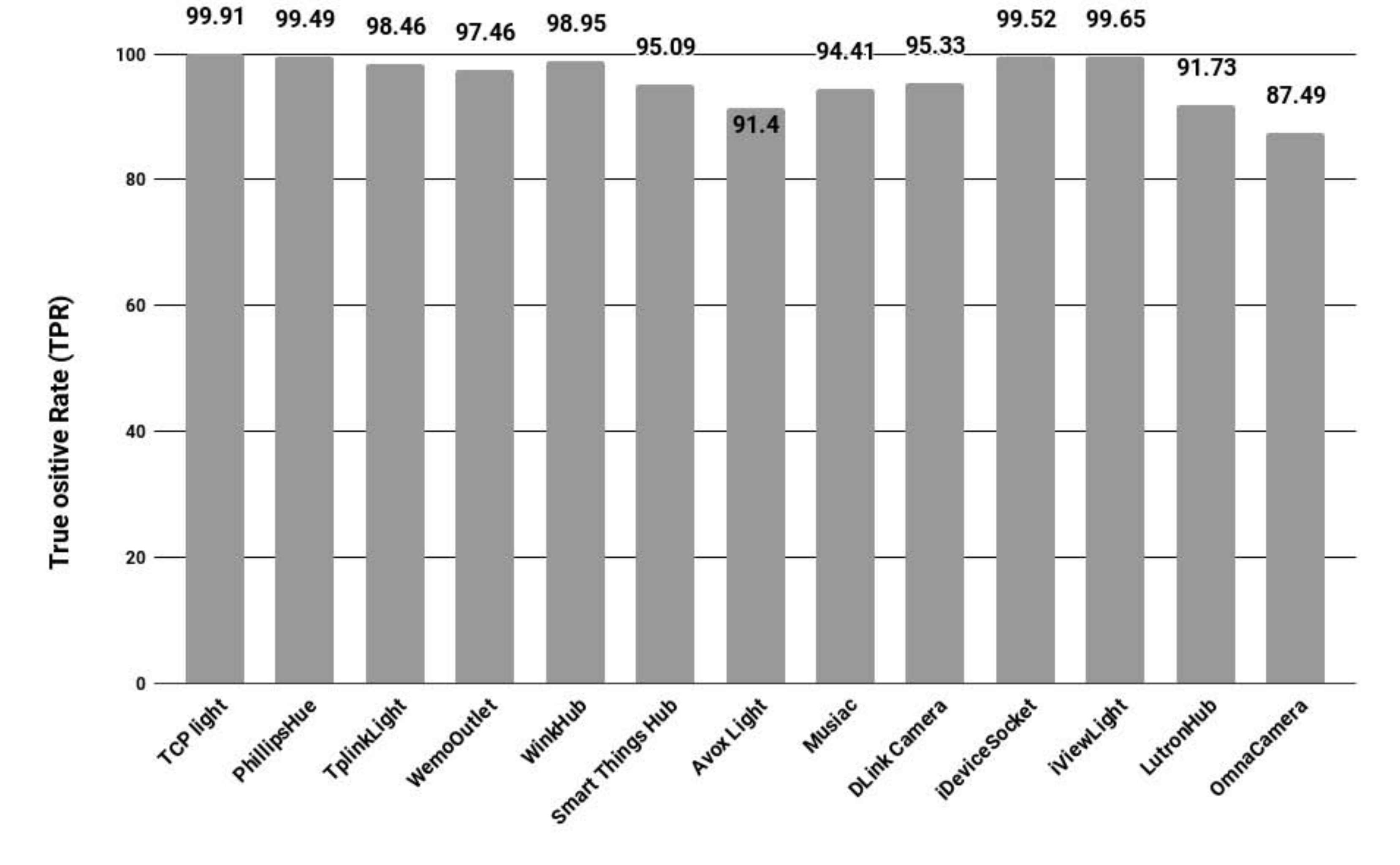}
  }
  \hspace*{-0.1in}
%\subfigure[3 Features]{\label{fig:3-ci}
 % \includegraphics[width=2in, height=1.8in]{CI-TPR-03.pdf}
  %}
\vspace*{-0.2in}
\caption{Device-type Identification Rate: (a) With all 20 features (b) Without entropy}\label{fig:tpr}
}
\vspace*{-0.1in}
\end{figure*}

\begin{figure*}[htbp]
%\centering
%\hspace*{-0.2in}
\parbox{6in}{
\subfigure[20 Features]{\label{fig:20-acc-ci}
  \includegraphics[width=3in, height=2in]{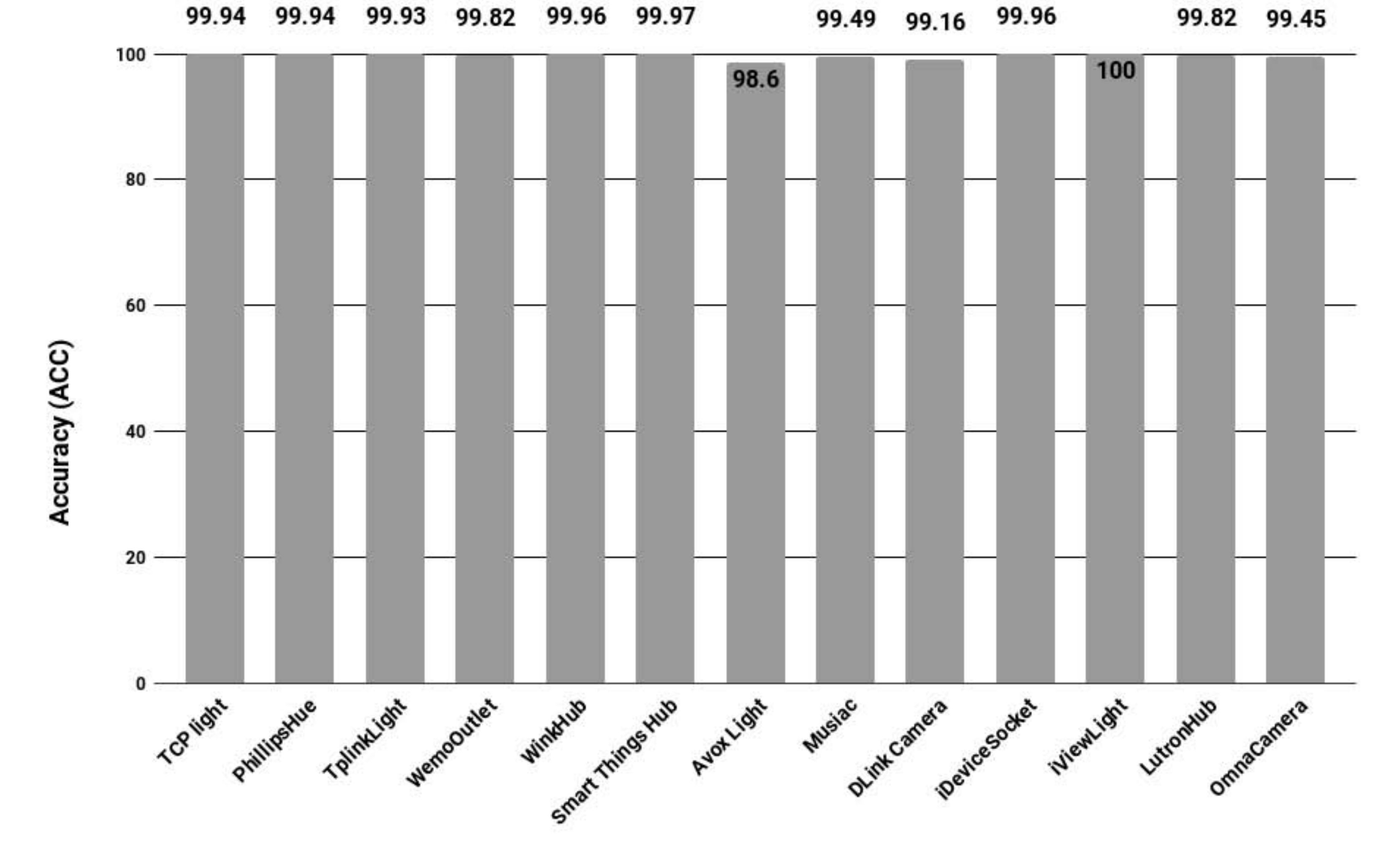}
  }
\hspace*{-0.1in}
\subfigure[19 Features]{\label{fig:19-acc-ci}
  \includegraphics[width=3in, height=2in]{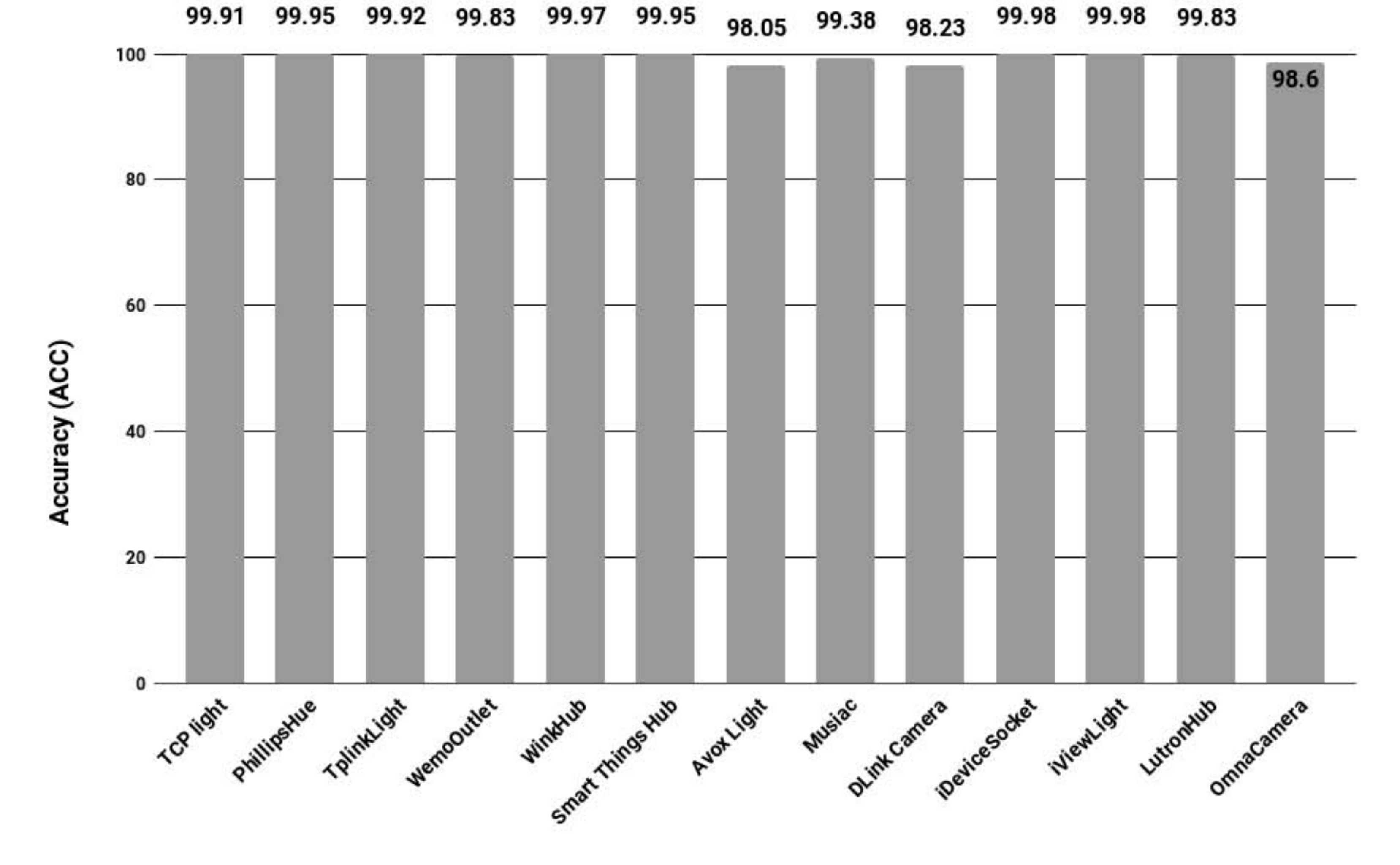}
  }
  \hspace*{-0.1in}
%\subfigure[3 Features]{\label{fig:3-acc-ci}
%  \includegraphics[width=2in, height=1.8in]{CI-ACC-03.pdf}
%  }
\vspace*{-0.2in}
\caption{Device-type Classification Accuracy: (a) With all 20 features (b) Without entropy}\label{fig:acc-ci}
}
\vspace*{-0.1in}
\end{figure*}

\begin{figure*}[htbp]
%\centering
%\hspace*{-0.2in}
\parbox{6in}{
\subfigure[20 Features]{\label{fig:20-tpr-cc}
  \includegraphics[width=3in, height=2in]{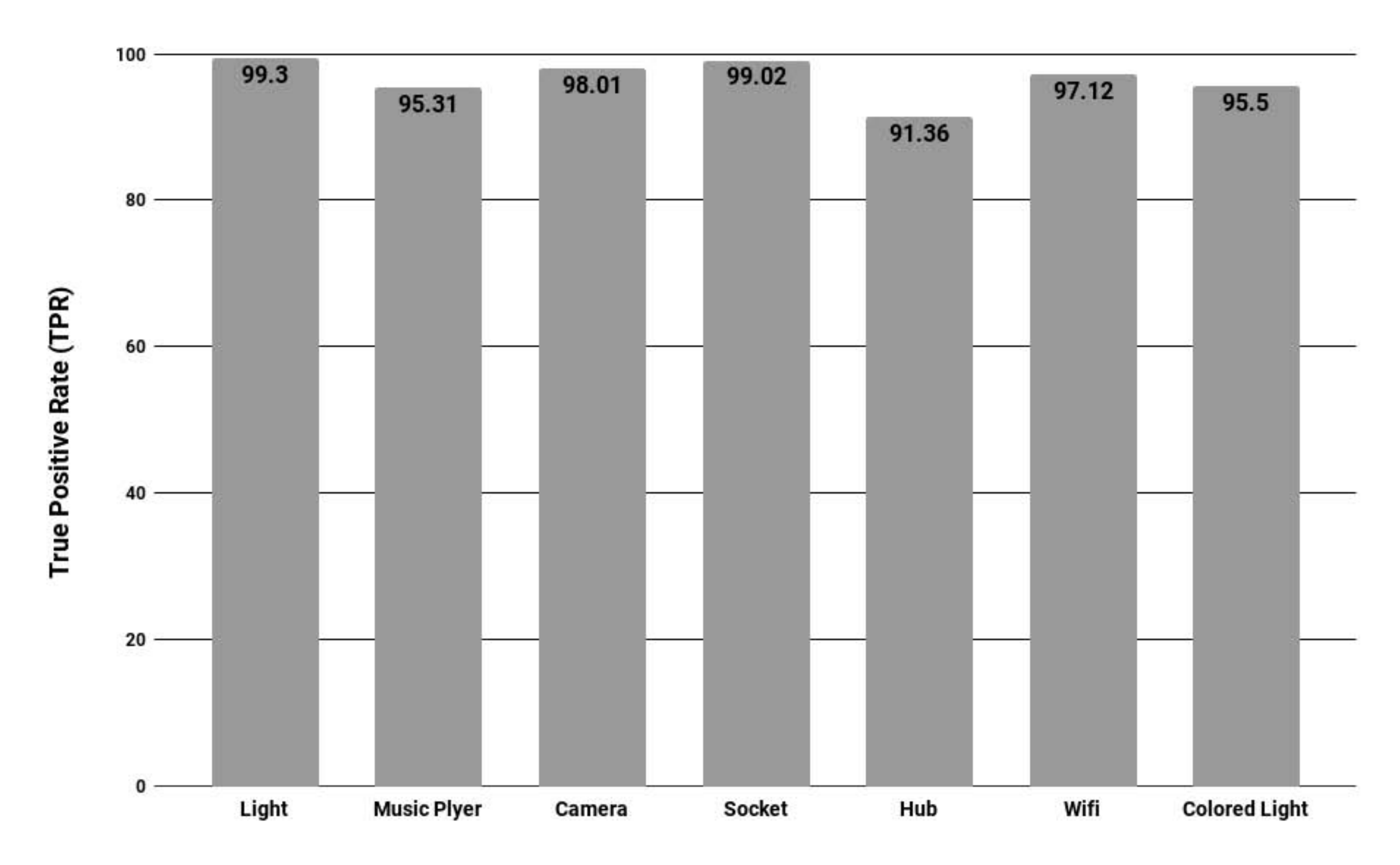}
  }
\hspace*{-0.1in}
\subfigure[19 Features]{\label{fig:19-tpr-cc}
  \includegraphics[width=3in, height=2in]{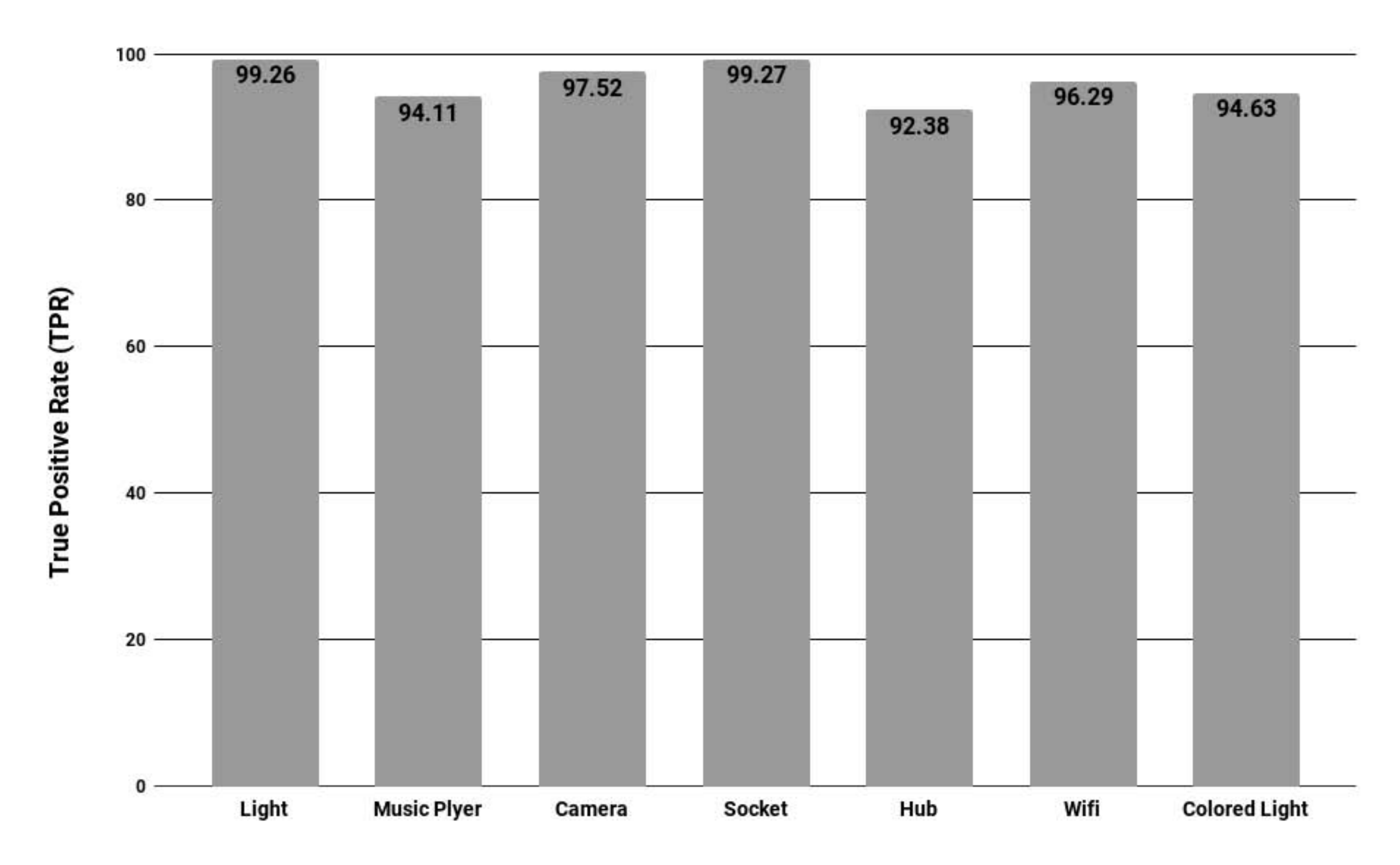}
  }
  \hspace*{-0.1in}
%\subfigure[3 Features]{\label{fig:3-CI}
%  \includegraphics[width=2in, height=1.8in]{CC-TPR-03.pdf}
%  }
\vspace*{-0.2in}
\caption{Device Category Identification Rate: (a) With all 20 features (b) Without entropy}\label{fig:tpr-cc}
}
\vspace*{-0.1in}
\end{figure*}
To generate a single data instance, based on the discussion in Section \ref{sec:dynamic} and Section \ref{sec:behavior}, we aggregated five consecutive packets into one feature vector.
The resulting data instances for each device are shown in Table \ref{tbl:data}.
\subsection{Machine Learning Classifier}\label{sec:learning}
\begin{figure*}[htbp]
%\centering
%\hspace*{-0.2in}
\parbox{6in}{
\subfigure[20 Features]{\label{fig:20-acc-cc}
  \includegraphics[width=3in, height=2in]{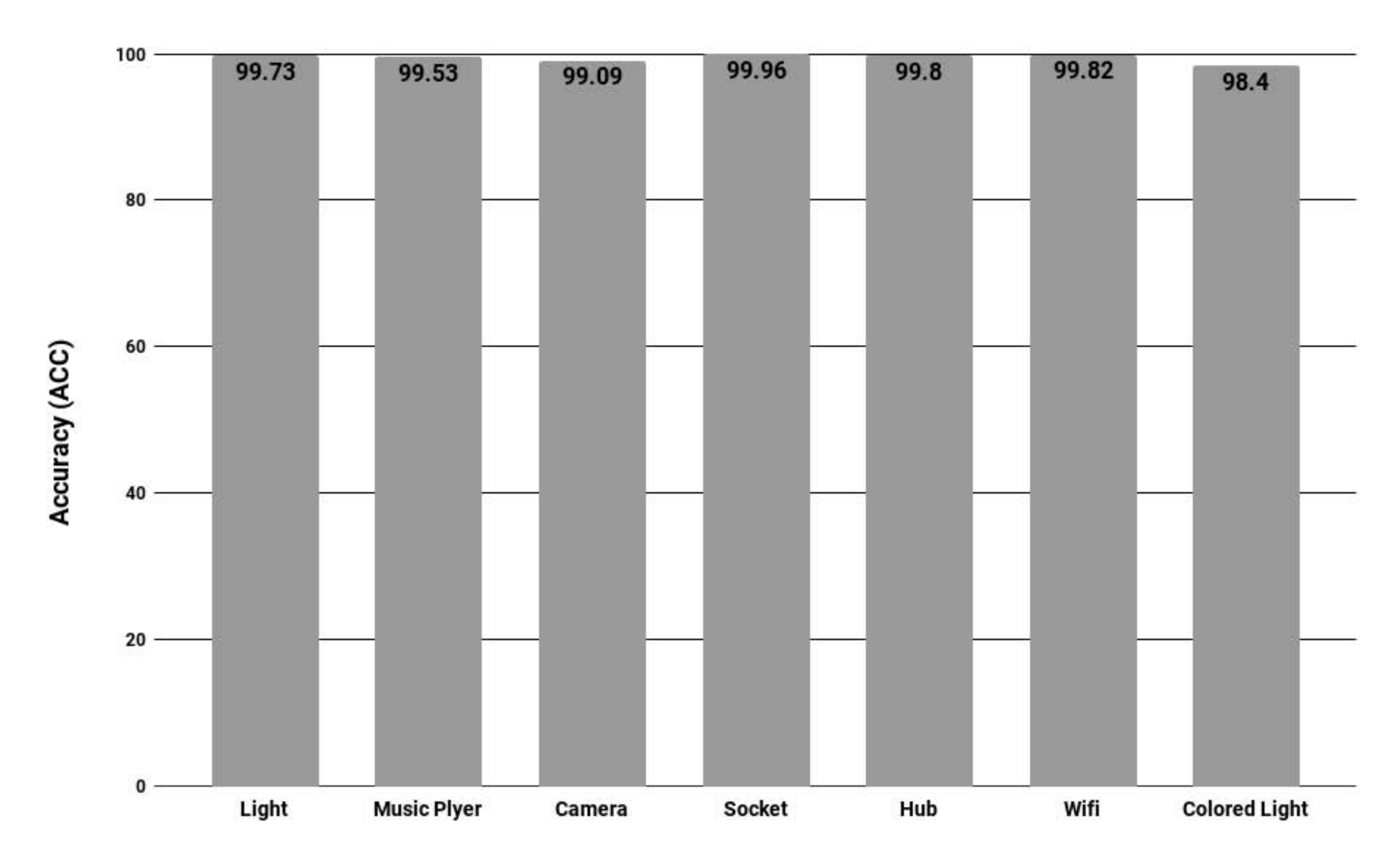}
  }
\hspace*{-0.1in}
\subfigure[19 Features]{\label{fig:19-acc-cc}
  \includegraphics[width=3in, height=2in]{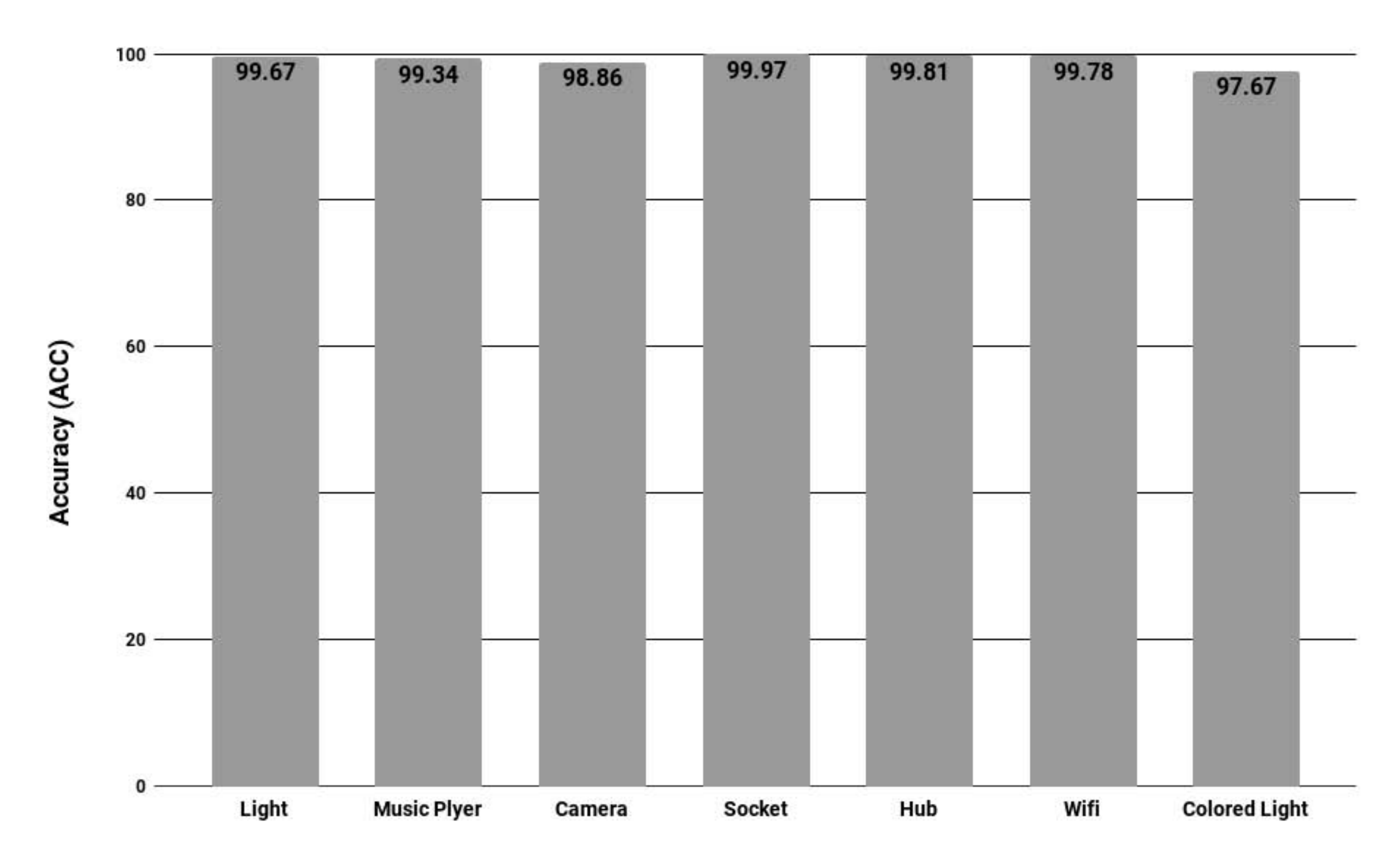}
  }
  \hspace*{-0.1in}
%\subfigure[3 Features]{\label{fig:3-CI}
%  \includegraphics[width=2in, height=1.8in]{CC-TPR-03.pdf}
%  }
\vspace*{-0.2in}
\caption{Device Category Classification Accuracy: (a) With all 20 features (b) Without entropy}\label{fig:acc-cc}
}
\vspace*{-0.1in}
\end{figure*}

We used several classifiers available in Scikit-learn tool \cite{pedregosa2011scikit} such as k-nearest-neighbors, Decision trees, Gradient boosting and Majority voting.
We describe Gradient boosting here, as this classifier gave consistently good results across all the experiments.
Gradient boosting \cite{mason2000boosting,friedman2002stochastic} is a gradient descent based learning approach that produces a prediction model as an ensemble of weak prediction models.
The learning starts with a "weak" model, typically Gradient Boost Regression Tree (GBRT) that tries to learn the data space and is iteratively improved by the next model that reduces the error of the previous model.
The goal of gradient boosting is to combine weak learning models into a single strong model as shown:
$F(x)=\sum^M_{m=1} \gamma_m h_m (x)$ where $F(x)$ is the closed function of the learning model.
Typically, $h_m$ is GBRT of fixed depth, which is iteratively improved over $M$ trials and $\gamma_m$ is the regression parameter for that particular iteration.
At each step, the model is improved as follows:
$$F_{m+1}(x)=F_m(x) + \gamma_{m+1} h_{m+1} (x)$$
The $h_{m+1}$ is chosen to minimize the loss function $L$ in the current model's fitting of a data point $x_i$: $F_m(x_i)$ as shown: %
$$F_{m+1}(x)=F_m(x)+ \textsf{arg} \underset{h}{\textsf{min}} \sum^n_{i=1} L(y_i, F_m(x_i)+h(x))$$
For implementation we used Scikit-learn library \cite{pedregosa2011scikit} and we set the tool-kit specific parameters as follows: $ n\_estimators=100$, which denotes the number of weak learners, and the maximum depth of each tree is controlled by $max\_depth$ parameter.
We set the $ learning\_rate=1.0 $ and $ max\_depth=1$.
\subsection{Evaluation Metrics}
We use the following metrics to evaluate our approach.
\begin{itemize}
\item {\em Identification Rate.}    \ This is essentially the true positive rate (TPR) of the classifier, \ie, the number of correctly identified (classified) data points of a given device-type from among the actual number of available data points of this device type.
\item {\em Accuracy.}   \ This is the ratio of the number of times the classifier correctly labels a data point to its correct class to the total number of available data points.    
\end{itemize}
We also evaluated other standard metrics like True negative rate (TNR) and Positive Predictive Value (PPV), but did not report them as we implemented one classifier per device-type.
These metrics are more meaningful if there was a single classifier with multiple classes.
Our classification methodology is "one-vs-all", \ie, the tested device data is labeled as "1" and the other data are labeled as "-1", which could be any device-type different from the device being tested.
\subsection{Device-type Fingerprinting}\label{sec:difp}
In the first experiment, we experimented the accuracy of the classifier for device-type fingerprinting.
The experiment is as  follows.
For testing against a given device, we treated the class label of the device to be $1$ and the rest of the $13$ devices data as $-1$.
Therefore, our learning model created an imbalance in the data wherein the positive labels were less than $10\%$ of the total data set.
We used five-fold cross-validation to avoid issues of over-fitting and to test the robustness of the classifier learning against unknown data instance classification.
We performed the experiment under three different conditions.

In the first variation, we included all the $20$ features we described in Section \ref{sec:features}.
As shown in Figure \ref{fig:20-ci}, this experiment achieved an high identification rate of $99\%$ for most of the devices except Lutron Hub, which did not have many data instances available.

In the second variation, we removed entropy from the feature set and experimented with the remaining $19$ features.
The reason for this experiment is to show that our approach works even when data is encrypted.
By not considering entropy, we eliminate any possibility that the results are biased due to the payload content.
In this experiment,  as shown in Figure \ref{fig:19-ci}, we achieved nearly identical results with a slight drop of $1$-$2\%$ overall.

We also performed a third variation, which we do not report here, where we experimented only with the $3$ payload based features.
The reason for this experiment was to see how well these features can perform in isolation.
The results were highly encouraging and there was only a slight drop from the experiment that used all $20$ features.
\begin{figure*}[htbp]
%\centering
%\hspace*{-0.2in}
\parbox{6in}{
\subfigure[20 Features]{\label{fig:20-cd-tpr}
  \includegraphics[width=3in, height=2in]{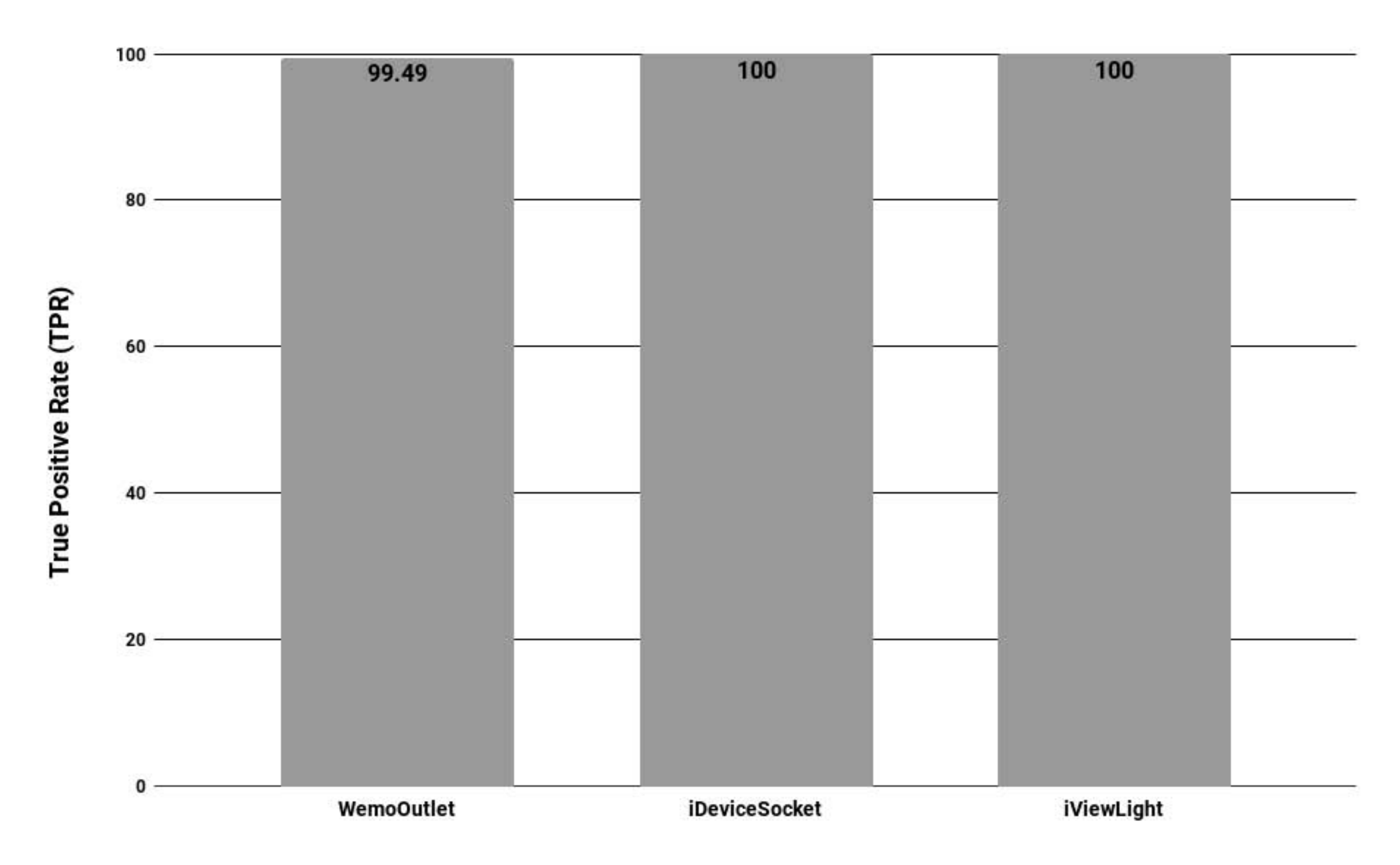}
  }
%\hspace*{-0.1in}
%
\subfigure[19 Features]{\label{fig:19-cd-tpr}
  \includegraphics[width=3in, height=2in]{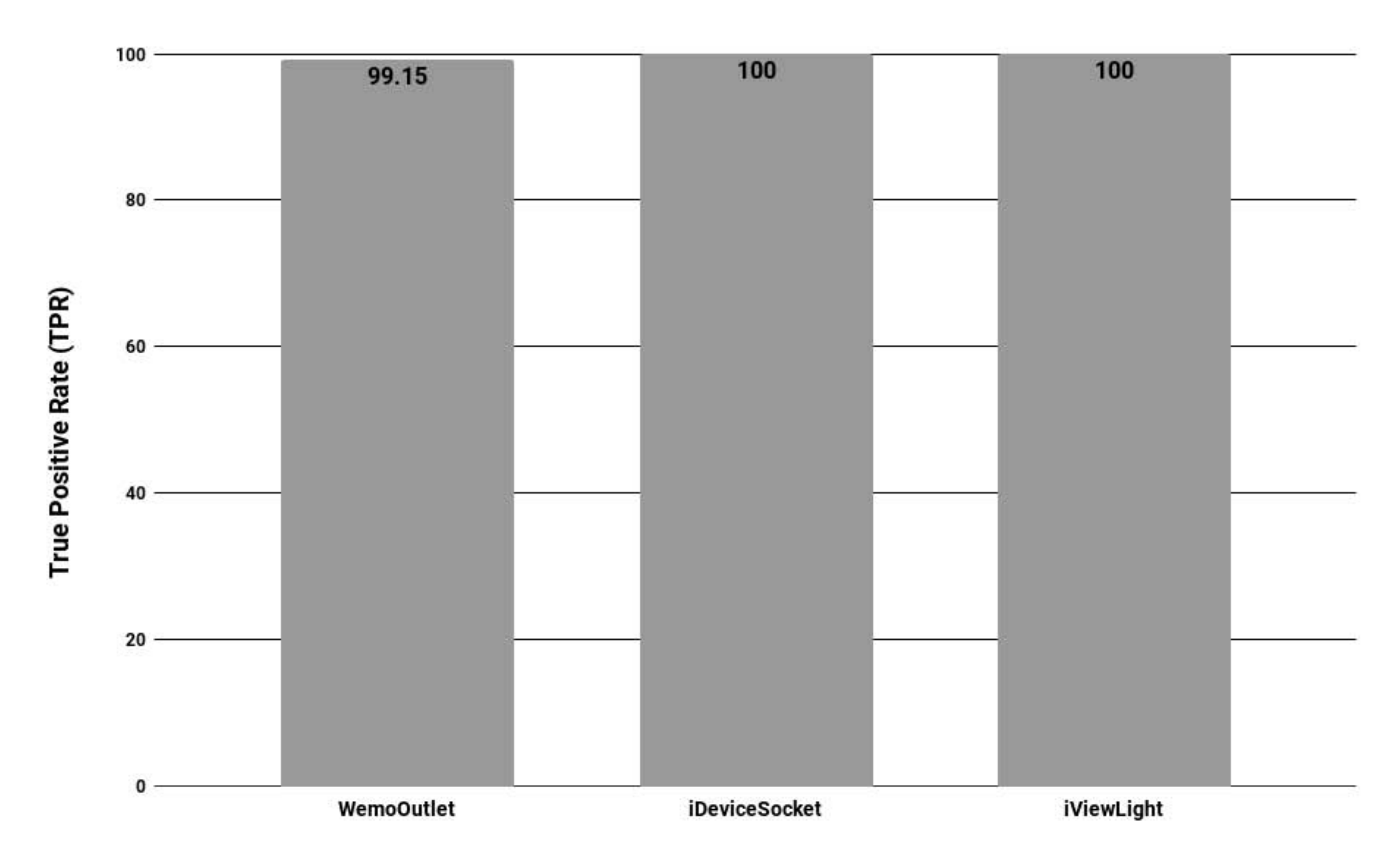}
  }
  \hspace*{-0.1in}
%\subfigure[3 Features]{\label{fig:3-CI}
 % \includegraphics[width=2in, height=1.8in]{CI-TPR-03.pdf}
 % }
\vspace*{-0.2in}
\caption{Device Recognition Rate: (a) With all 20 features (b) Without entropy}\label{fig:tpr-cd}
}
\vspace*{-0.1in}
\end{figure*}

\begin{figure*}[htbp]
%\centering
%\hspace*{-0.2in}
\parbox{6in}{
\subfigure[20 Features]{\label{fig:20-dd-tpr}
  \includegraphics[width=3in, height=2in]{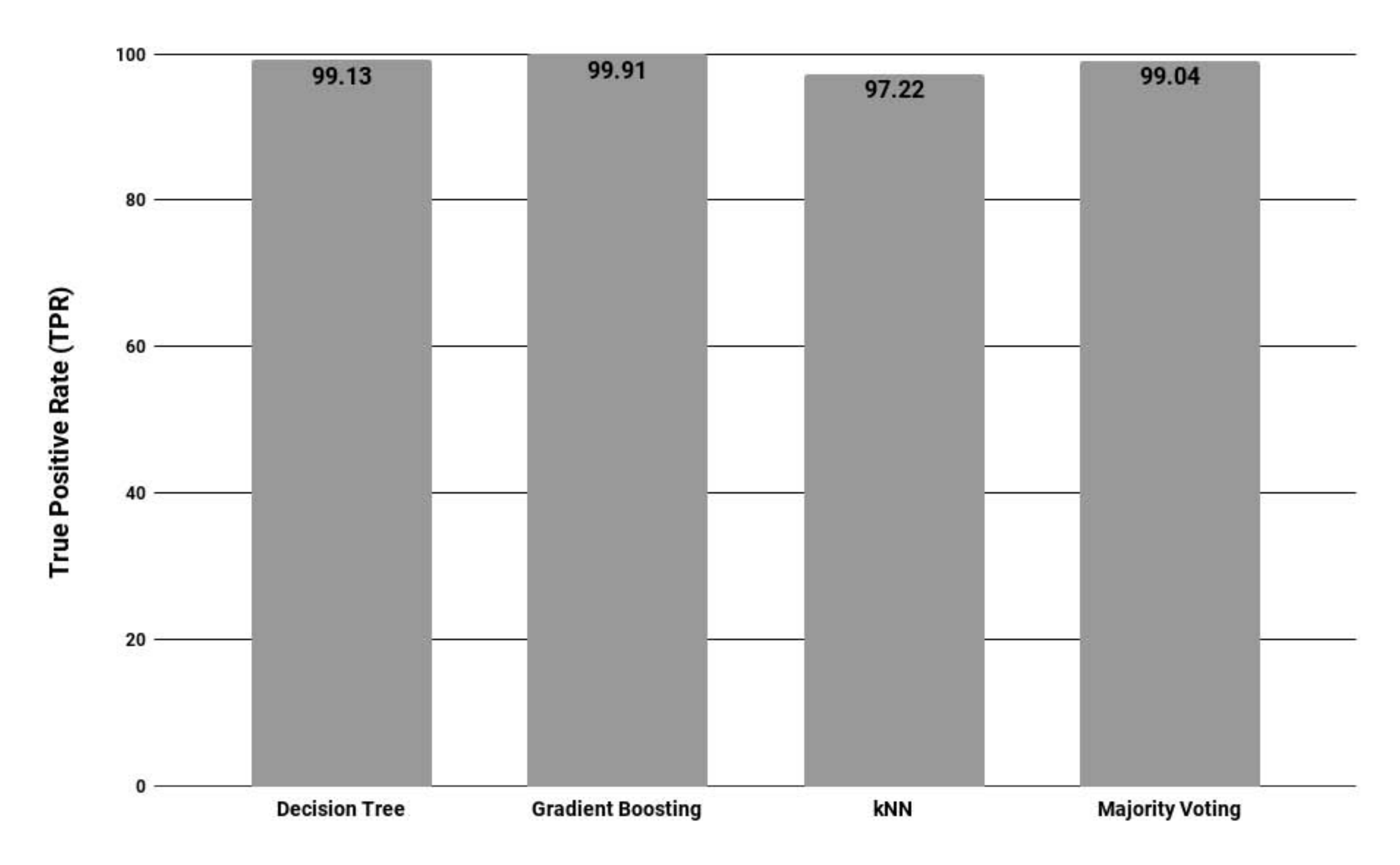}
  }
%\hspace*{-0.1in}
%
\subfigure[20 Features]{\label{fig:19-cd-tpr}
  \includegraphics[width=3in, height=2in]{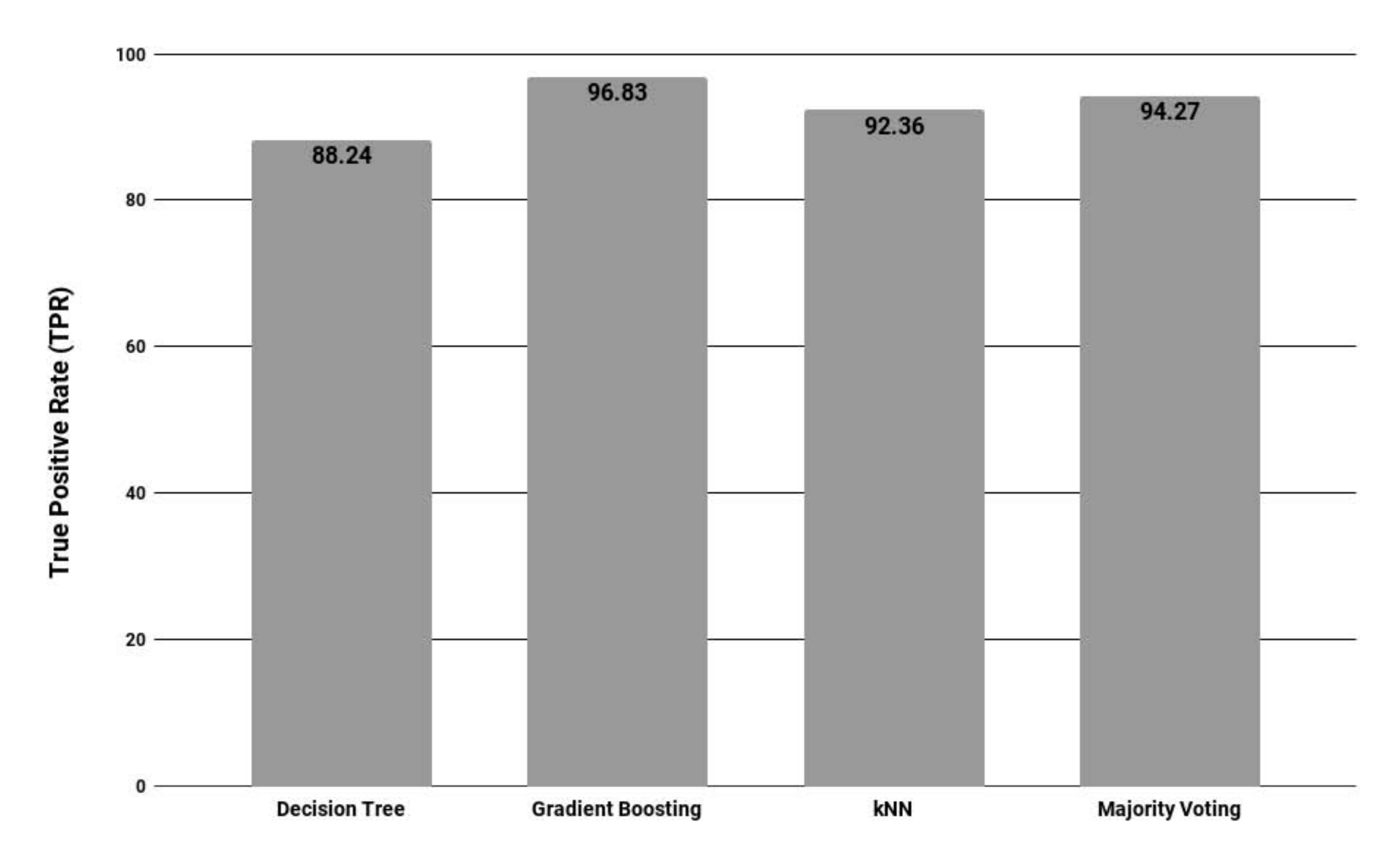}
  }
  \hspace*{-0.1in}
%\subfigure[3 Features]{\label{fig:3-CI}
 % \includegraphics[width=2in, height=1.8in]{CI-TPR-03.pdf}
 % }
\vspace*{-0.2in}
\caption{Device-type Identification Rate: (a) TCP Light (b) D-Link Camera}\label{fig:tpr-dd}
}
\vspace*{-0.1in}
\end{figure*}

\begin{figure*}[htbp]
%\centering
%\hspace*{-0.2in}
\parbox{6in}{
\subfigure[20 Features]{\label{fig:20-dd-acc}
  \includegraphics[width=3in, height=2in]{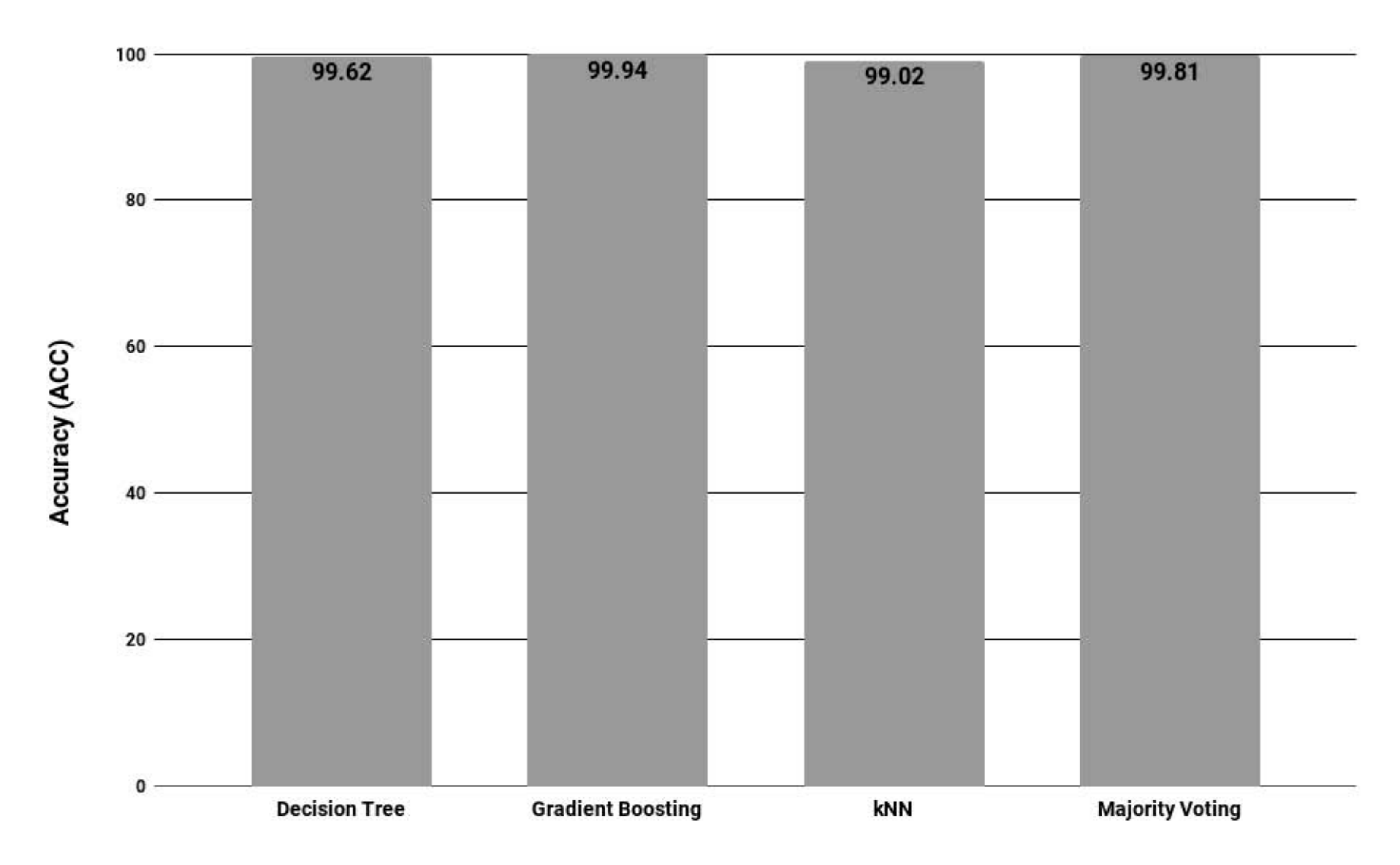}
  }
%\hspace*{-0.1in}
%
\subfigure[20 Features]{\label{fig:19-dd-acc}
  \includegraphics[width=3in, height=2in]{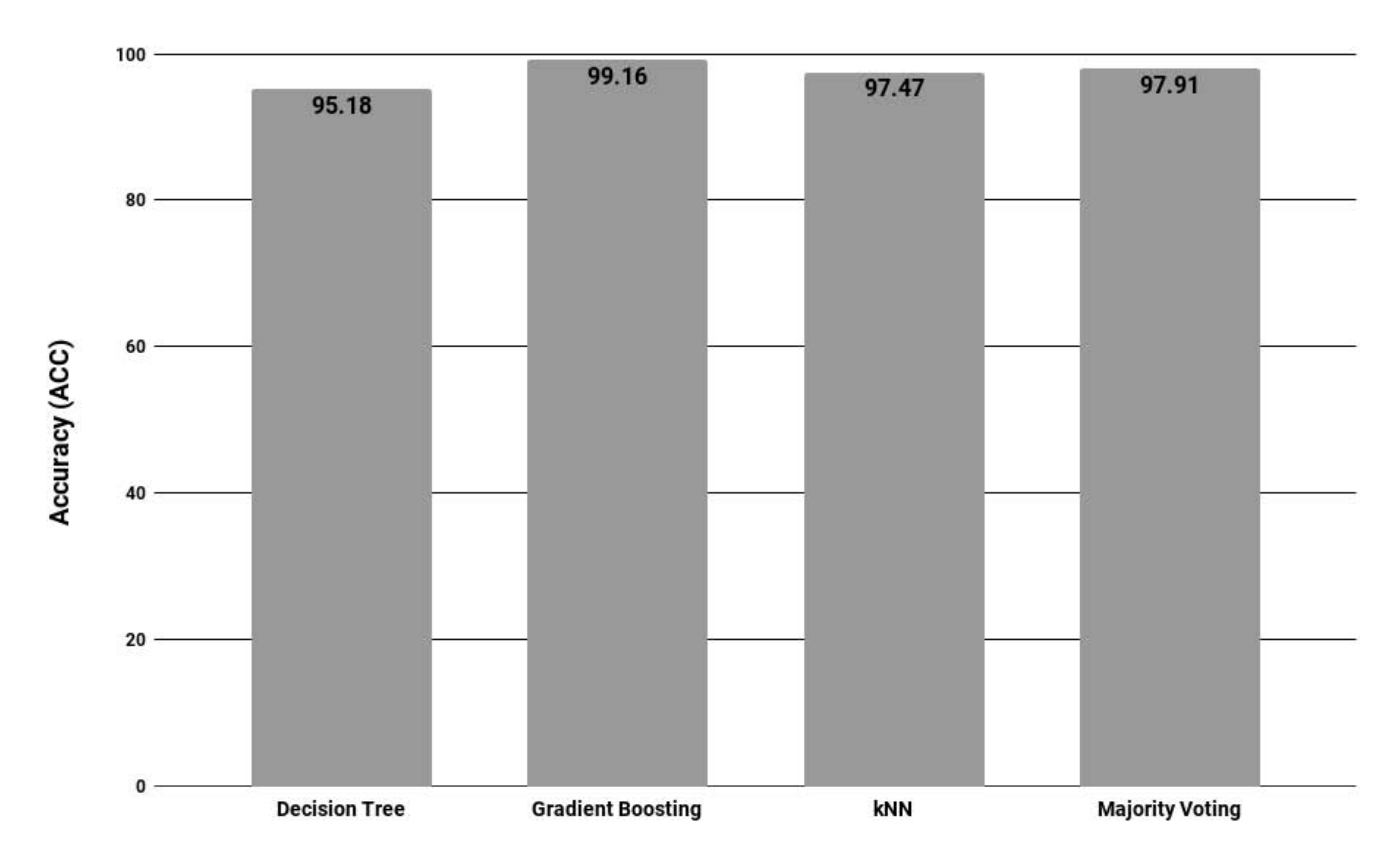}
  }
  \hspace*{-0.1in}
%\subfigure[3 Features]{\label{fig:3-CI}
 % \includegraphics[width=2in, height=1.8in]{CI-TPR-03.pdf}
 % }
\vspace*{-0.2in}
\caption{Device-type Classification Accuracy: (a) TCP Light (b) D-Link Camera}\label{fig:acc-dd}
}
\vspace*{-0.1in}
\end{figure*}

In Figure \ref{fig:acc-ci}, we show the average accuracy across the devices, which is consistently above $99\%$.
This result is very significant given the skewed nature of our data set, most of the times the classifier was correctly able to recognize that a particular data instance did not correspond to the target class label.
In real-world networks, this is the most likely situation for fingerprinting and it is essential that the classifier does not generate too many false positives.
\begin{table}[hbp]
\small
\centering
\vspace*{-0.1in}
\caption{Device category Data}
\vspace*{-0.1in}
\label{tbl:crossdata}
\begin{tabular}{|l|c|}
\hline
\multicolumn{1}{|l|}{\textbf{Device category}} & \textbf{Data Instances} \\\hline
1: Light                            &  2422 \\\hline
2: Music Player                          &  1003\\\hline
3:	Camera                   &  3063 \\\hline
4: 	Socket                               & 415 \\\hline
5: Hub                           &   210 \\\hline
6: 	Outlet                        &  592 \\\hline
7: Colored Light                          &  3090  \\\hline
\end{tabular}
\end{table}
\subsection{Device category Fingerprinting}\label{sec:dcfp}
In this experiment, we explored the capability of our model in classifying devices into device-types.
For this purpose, we generated a data set from the original data set by grouping devices into device-types, \eg, light bulbs.
The data set is shown in Table \ref{tbl:crossdata}, which consists of categories described in Table \ref{tbl:data}.
\begin{table}[hbp]
\small
\centering
\caption{Device-instance Data}
\vspace*{-0.1in}
\label{tbl:dd-data}
\begin{tabular}{|l|c|}
\hline
\multicolumn{1}{|l|}{\textbf{Device Label: Device}} & \textbf{Data Instances} \\\hline
5 :iDeviceSockett                            &  415 \\\hline
6: iViewLight                          &  571\\\hline
12:	WemoOutlet               & 592 \\\hline
\end{tabular}
\end{table}

The device category identification rate ranged from $93$-$99\%$ across the different device-types as shown in Figure \ref{fig:tpr-cc}.
This result is very significant as it demonstrates the feasibility of categorizing devices into common device-types.
Our work is the first to report this kind of result.
Furthermore, as shown in Figure \ref{fig:acc-cc}, the average accuracy for this experiment is in the range of $97$-$99\%$, which shows the robustness of the classifier even against noisy data.
\subsection{Cross-instance Recognition}\label{sec:cifp}
In this experiment, we collected data from different device instances of Wemo Outlet, iDevice Socket and iView light, shown in Table \ref{tbl:dd-data}.
The idea was to check how well the learning model could recognize different device instances.
The training set used for this experiment was from Table \ref{tbl:data}, as would be required in a live scenario.
The experiments reported a high recognition rate of $99.7$-$100\%$, indicating that the classifier was very successful in matching instances against previously stored profiles of the device type.
\subsection{Performance Across Multiple Classifiers}\label{sec:classifiers}
All the above experiments were repeated across multiple machine learning classifiers like: k-nearest-neighbors (kNN), Decision tree and Majority voting.
The classifiers still reported a high identification rate (true positive rate) ranging from  $88$-$99\%$, shown in Figure \ref{fig:tpr-dd}, and an average accuracy ranging from $95$-$99\%$, shown in Figure \ref{fig:acc-dd}.
We show the results for device type experiment on two selected devices: TCP Light and D-Link camera, for $20$ features.
\section{Conclusion}\label{sec:conclusion}\label{sec:conclusion}
In conclusion, we re-affirm that the problem of IoT device  fingerprinting is very important in the context of security.
The identification of IoT device types is a strong step towards identifying IoT device instances, which will be useful in establishing strong authentication of a device.
The existing IoT devices have too much variation in protocols/functionality and it is difficult to come out with one general approach for fingerprinting.
However, as our methodology showed, it is possible to fingerprint device types with high accuracy.
The  high accuracy reported by our experiments show that it is possible to reduce false positives during device fingerprinting even in the presence of several other devices.
Fingerprinting categories of devices is an entirely different challenge and we demonstrated some promising results in this direction.
Our work is the first to report such cross-category identification of devices.
There are many open questions remaining in fingerprinting and this will continue to be an interesting research area for the IoT domain for quite some time.
One question is how to leverage device fingerprinting towards the development of strong authentication schemes for IoT devices that are difficult to clone.
Such schemes should not require manufacturers to make changes to their device architectures or enforce unreasonable computational overhead on the devices themselves.

\bibliographystyle{IEEEtran}
\bibliography{network}

\end{document}